\documentclass[%
 reprint,
superscriptaddress,
bibnotes,
amsmath,amssymb,
aps,
pra,
floatfix,
]{revtex4-2}

\usepackage{graphicx}% Include figure files
\usepackage{dcolumn}% Align table columns on decimal point
\usepackage{bm}% bold math
\usepackage{hyperref}% add hypertext capabilities
\usepackage{chemformula} % Formula subscripts using \ch{}
\usepackage{physics}
\usepackage{upgreek}
\usepackage{cleveref}

\graphicspath{{./Figures/}}
\newcommand{\etal}{\textit{et al. }}
\newcommand{\tu}[1]{\textup{#1}}

\newcommand{\pdvn}[1]{\frac{\partial {#1}}{\partial n}}

\newcommand{\RNum}[1]{\uppercase\expandafter{\romannumeral #1\relax}}
\newcommand{\SI}[1]{\textit{Supplementary section {#1}}}

\renewcommand{\figurename}{Fig.}
\makeatletter
\renewcommand*{\fnum@figure}{{\normalfont\bfseries \figurename~\thefigure}}
\makeatother
\renewcommand{\tablename}{Table.}
\makeatletter
\renewcommand*{\fnum@table}{{\normalfont\bfseries \tablename~\thetable}}
\makeatother

\Crefformat{equation}{Eq.~#2#1#3}
\Crefrangeformat{equation}{Eqs.~#3#1#4 to~#5#2#6}
\Crefmultiformat{equation}{Eqs.~#2#1#3}{ and~#2#1#3}{, #2#1#3}{ and~#2#1#3}
\Crefformat{figure}{Fig.~#2#1#3}
\Crefrangeformat{figure}{Figs.~#3#1#4 to~#5#2#6}
\Crefmultiformat{figure}{Figs.~#2#1#3}{ and~#2#1#3}{, #2#1#3}{ and~#2#1#3}
\Crefformat{table}{Table.~#2#1#3}
\Crefrangeformat{table}{Tables.~#3#1#4 to~#5#2#6}
\Crefmultiformat{table}{Tables.~#2#1#3}{ and~#2#1#3}{, #2#1#3}{ and~#2#1#3}

\begin{document}

%\preprint{APS/123-QED}

\title{Crystal Nucleation Driven by Microjets from Hydrodynamically Interacting Thermocavitation Bubbles}% Force line breaks with \\
%\thanks{denotes equal contribution}
% Laser induced crystal nucleation facilitated bu microjets emerging from hydrodynamical interaction of thermocaviation bubble paied under confinement

\author{Nagaraj Nagalingam}
\affiliation{Process \& Energy Department, Delft University of Technology, Leeghwaterstraat 39, 2628 CB Delft, Netherlands}
\author{Adithya Vignesh}
\affiliation{Process \& Energy Department, Delft University of Technology, Leeghwaterstraat 39, 2628 CB Delft, Netherlands}
\author{Vikram Korede}
\affiliation{Process \& Energy Department, Delft University of Technology, Leeghwaterstraat 39, 2628 CB Delft, Netherlands}
\author{Herman J.M. Kramer}
\affiliation{Process \& Energy Department, Delft University of Technology, Leeghwaterstraat 39, 2628 CB Delft, Netherlands}
\author{Peter J.T. Verheijen}
\affiliation{Biotechnology Department, Delft University of Technology, Van der Maasweg 9, 2629 HZ Delft, Netherlands}
\author{Jan Groenewold}
\affiliation{Debye Institute for Nanomaterials Science, Utrecht University, Padualaan 8, 3584 CH Utrecht, Netherlands}
\author{Remco Hartkamp}
\affiliation{Process \& Energy Department, Delft University of Technology, Leeghwaterstraat 39, 2628 CB Delft, Netherlands}
\author{Johan T. Padding}
\affiliation{Process \& Energy Department, Delft University of Technology, Leeghwaterstraat 39, 2628 CB Delft, Netherlands}
\author{Hüseyin Burak Eral}
\email{h.b.eral@tudelft.nl}
\affiliation{Process \& Energy Department, Delft University of Technology, Leeghwaterstraat 39, 2628 CB Delft, Netherlands} 

\date{\today}% It is always \today, today,
\begin{abstract}
Primary nucleation predominantly dictates the physicochemical, optical and mechanical properties of crystals emerging from solution. Here, we reveal a previously unrecognized nucleation mechanism arising from bubble-bubble hydrodynamic interactions using laser-induced thermocavitation bubble pairs. Under confinement, antiphase generation of two bubbles produces a micron-scale, high-speed liquid jet. Upon impact with the surrounding supersaturated liquid, this microjet generates intense localized shear, thereby inducing crystallization of the model inorganic salt, potassium permanganate. Under identical laser energies and supersaturation levels, crystallization was not observed in single-bubble experiments, in which no microjet was present. We show that the crystallization probability increases with microjet velocity. Boundary integral simulations reveal how experimental parameters such as confinement, bubble spacing, and laser energy govern microjet formation and velocity. We quantify the relationship between jet velocity and the observed enhancement in crystal nucleation kinetics by incorporating shear into the kinetic prefactor in classical nucleation theory. By combining experiments, theory, and simulations, we establish how microjets emerging from bubble-bubble hydrodynamic interactions may influence crystallization kinetics, opening new opportunities for the manufacture of crystalline materials.
\end{abstract}

\maketitle

% Abstract
%\onecolumngrid 

% Introduction
%\clearpage
%\twocolumngrid

%-
%\clearpage \pagebreak

% Here, we demonstrate that confined pairs of laser-induced bubbles can cooperatively trigger crystallization through the generation of high-speed micro-jets

\textit{Significance:} From hydrothermal mineralization in deep-sea vents to industrial manufacturing, crystallization often occurs under flow. However, despite the pervasive influence of hydrodynamics, its role in regulating nucleation pathways remains largely overlooked. We demonstrate a mechanistically distinct route to nucleation via microjet induced shear generated by cooperative hydrodynamic interactions between thermocavitation bubble pairs. By engineering bubble-bubble interactions, we generate microjets that produce highly localized shear rates of  $O(10^5\,\mathrm{s}^{-1})$ that persist for only a few microseconds but substantially accelerate crystal nucleation from solution. The proposed crystallization pathway opens new avenues in flow-assisted crystallization with applications spanning from pharmaceuticals to organic semiconductors.

Nucleation from supersaturated solution is of fundamental interest in understanding how nature self-organizes seemingly uncorrelated molecules into complex crystalline structures \cite{10.1126/science.1243022,10.1126/sciadv.aav7399,10.1126/science.aaa6760}. Moreover, controlling nucleation has practical relevance for applications ranging from the manufacturing of functional materials that outperform their natural counterparts to understanding the mineral morphogenesis and the origin of homochirality \cite{10.1038/ncomms2490,10.1126/sciadv.abl9653,10.1142/5172,10.1126/science.1090163,10.1038/s41586-022-04561-z}. However, nucleation, growth and other crystallization phenomena that often occur in dynamic multiphase environments in the presence of flows render nucleation control challenging. The use of light has emerged as a promising non-invasive approach due to its ability to spatiotemporally control the crystal nucleation and therefore the properties of emerging crystals \cite{10.1021/acs.cgd.2c01526,10.1021/acs.cgd.3c00865,10.1021/acs.cgd.4c01042,10.1073/pnas.2419638122,10.1021/jacs.1c11154}. Specific to pulsed lasers, generally having femto- to nanosecond pulse width, the formation of crystals is associated with vapor (cavitation) bubbles formed due to the interaction of laser light with the solution \cite{10.1103/PhysRevLett.131.124001,10.1016/j.jphotochemrev.2022.100530}. Despite recent progress, predictive control over crystal quality parameters such as size, purity, (pseudo)polymorphic form and morphology along with nucleation rate remains elusive \cite{10.1021/acs.cgd.2c01526,10.1016/j.jphotochemrev.2022.100530}. 
%Therefore, systematic and concomitant studies employing experiments and theory/simulations are necessary to optimize the laser-induced crystallization technique.  HBE: You are not doing this so no need for this sentence

Although considerable effort has focused on controlling crystal quality parameters through thermodynamic variables, flow-induced hydrodynamic phenomena offer new opportunities in crystallization \cite{10.1039/d4cs00148f,10.1103/PhysRevE.93.042803, 10.1103/8ghn-yrlr}. Among these hydrodynamic phenomena, cavitation bubbles are ubiquitous: as they localize high-velocity flows and interact with each other and confining surfaces through the surrounding liquid \cite{10.1073/pnas.2503033122, 10.1016/j.ultsonch.2010.11.016}. However, the implications of these interactions for nucleation from solution remain largely underexplored \cite{10.1016/j.seppur.2026.139442,10.1021/acs.cgd.3c00232}. The cavitation of a single bubble, inducing strong localized flows and corresponding stresses, has been utilized in microfluidics \cite{10.1007/978-981-287-278-4_6,10.1038/s41598-024-68971-x} and biomedical applications \cite{10.1126/science.169.3948.869,10.1073/pnas.1208198109}. As long as the bubble is sufficiently far from walls, free surfaces and other bubbles, the bubble will retain a spherical shape throughout its lifetime \cite{10.1073/pnas.1805912115}. However, asymmetries in the bubble's environment can lead to anisotropic bubble dynamics, and even the formation of microjets \cite{10.1146/annurev.fl.19.010187.000531}. These microjets of fast-moving liquid form due to the impulse generated as a result of the bubble's interaction with its surroundings. Jetting consequently leads to transport of fluid and species away from the location of bubble inception. Common methods for induced microjets include the generation of bubbles near rigid boundaries, near free surfaces, or near other bubbles \cite{10.1017/jfm.2016.463,10.1103/PhysRevLett.105.078101,10.1073/pnas.2301956120}. The jet emerging from bubble formation close to a rigid boundary can often cause surface damage \cite{10.1016/j.ultsonch.2015.04.006,10.1063/1.2337506}, which is therefore considered undesired in certain applications. With free interfaces hardly present in continuously functioning synthetic or physiological systems, bubble-bubble interactions present a promising approach for jet creation. Thus, inducing jets through interaction between multiple laser-induced bubbles is a versatile and non-invasive approach to controlling jet properties and has found applications in microrheology \cite{10.1103/PhysRevLett.104.014501}, single-cell poration \cite{10.1103/PhysRevLett.105.078101,10.1073/pnas.1518679112}, mechanotransduction \cite{10.1073/pnas.1713905115}, and particle manipulation \cite{10.1063/1.3511538}. 

% The interaction of a bubble with its surrounding is mediated via flow of the liquid surrounding the bubble \cite{10.1103/PhysRevLett.98.254501} and shockwaves generated during bubble formation and collapse \cite{10.1063/1.5060645}.

% \begin{widetext}
%     \begin{minipage}[t][][t]{\linewidth}
%         \begin{figure}[H]
%         	\centering
%         	\includegraphics[width=\linewidth]{Figure_1_1.pdf}
%         	\caption{Sketch of the experimental setup to generate a microbubble. The setup construction is detailed in our previous work \cite{10.1016/j.ohx.2023.e00415}. The green arrow indicates the direction of laser pulse.}
%         	\label{Figure_1}
%         \end{figure}
%     \end{minipage}
% \end{widetext}
\begin{figure*}[!t]
    \centering
    \includegraphics[width=\linewidth]{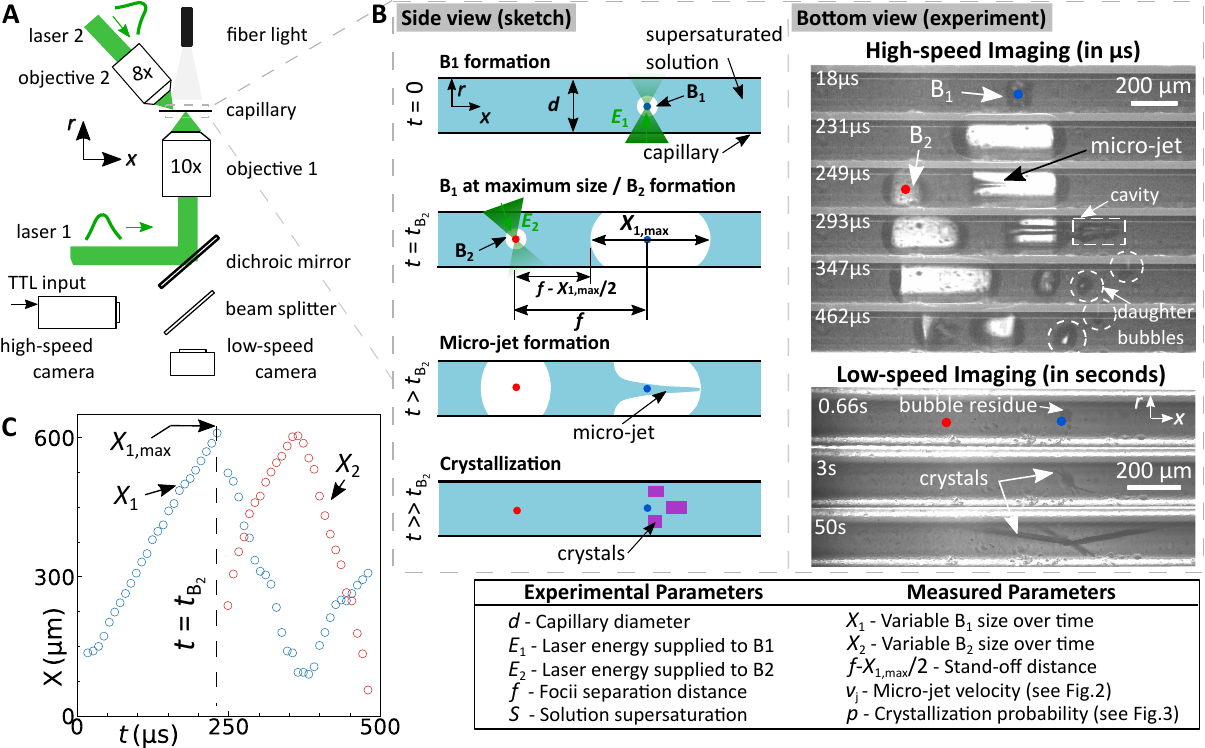}
    \caption{Experimental setup and method. ($A$) Optical paths of nanosecond laser pulses that travel through two separate objectives triggering thermocavitation bubbles. Two cameras, one high-speed and one low-speed, provide access to different timescales, O($\upmu$s) and O(s), respectively. ($B$) Illustration of thermocavitation bubbles interacting in a microcapillary. The first thermocavitation bubble ($\tu{B}_1$) is triggered at t=0. The second bubble $\tu{B}_2$ is triggered when $\tu{B}_1$ is at its maximum size, $X_\tu{1,max}$, thus making the bubbles antiphase. As $\tu{B}_2$ expands, a jet emerges within the $\tu{B}_1$ due to hydrodynamic interaction between bubbles. The distance of the center of $\tu{B}_2$ to nearest $\tu{B}_1$ interface at when $X_1=X_\tu{1,max}$ is $f-X_\tu{1,max}/2$, where $f$ is the distance between the laser foci. The daughter bubbles and cavity formed as a result of $\tu{B}_1-\tu{B}_2$ interaction is represented with white dotted lines using circles and rectangle, respectively. Long after the bubbles disappear, \ch{KMnO4} solute crystals emerge from solution. The crystals then grow in size over minutes. The experimental parameters prescribed prior to each experiment, together with the quantities measured during the experiment, are summarized in a table. ($C$) Measured dynamic bubble sizes from high-speed microscopy images illustrated in panel $B$. Bubble $\tu{B}_2$ is formed when $\tu{B}_1$ reaches its maximum size at $t=t_{\tu B_2}$. $X_1$ and $X_2$ correspond to the sizes of bubbles $\tu{B}_1$ and $\tu{B}_2$, respectively.}
    \label{Figure_1}
\end{figure*}

In this work, we show that hydrodynamic interactions between laser-induced thermocavitation bubbles can alter the kinetics of crystal nucleation in an aqueous supersaturated solution. Specifically, we identify a previously unexplored route to crystallization mediated by interacting bubble pairs confined within a microcapillary, see \Cref{Figure_1}. Under identical supersaturation conditions and laser energies, a single thermocavitation bubble does not induce crystallization within the experimental parameter range studied. In contrast, antiphase generation of two bubbles produces a micron-scale liquid jet traveling through the first bubble, driven by the expansion of the second bubble. Upon impact with the surrounding liquid, this microjet generates intense localized shear, whose magnitude scales with the velocity of the jet, leading to a substantial enhancement of the nucleation rate. In other words, the nucleation rate is observed to increase with increasing jet velocity in experiments. We first employ boundary integral simulations to elucidate how laser energy, bubble-bubble separation distance, and geometric confinement govern jet formation, morphology, and velocity. We then establish the relationship between jet velocity and crystal nucleation kinetics and interpret the observed behavior within the framework of classical nucleation theory. By integrating experiments, theory, and simulations, we demonstrate how bubble-bubble interactions under confinement can be engineered to control nucleation kinetics in solution.\\

\noindent\textbf{\large Results}\\
\noindent \textbf{Experiments}\\ 
\indent Bubble pairs are generated using two 532 nm pulsed lasers (pulse width 5-9 ns) synchronized through a delay generator. Vapor bubbles $\tu{B}_1$ and $\tu{B}_2$ are created by lasers 1 and 2, respectively, as illustrated in \Cref{Figure_1}. The dynamics of the bubbles and the resulting jet are governed by the absorbed laser energies, $E_1$ and $E_2$, which determine the maximum bubble sizes, $X_{\tu{1,max}}$ and $X_{\tu{2,max}}$, the focal separation distance $f$, the effective standoff distance $f-X_{\tu{1,max}}/2$, and the time delay between bubble initiation. The formation times of bubbles $\tu{B}_1$ and $\tu{B}_2$ are denoted by $t_{\tu B_1}$ and $t_{\tu B_2}$, respectively, with $t_{\tu B_1}=0$. The effective standoff distance, defined at $t=t_{\tu B_2}$, can be adjusted either by varying the focal separation $f$ or by modifying $X_{\tu{1,max}}$. The latter is controlled through the laser energy supplied to $\tu{B}_1$ ($E_1$) and the capillary diameter $d$ \cite{10.1038/s41598-024-68971-x}. Consequently, the experimentally prescribed parameters are the capillary diameter $d$, focal separation $f$, laser energies $E_1$ and $E_2$, and solution supersaturation $S$. The primary observables are the jet dynamics characterized by the jet velocity, and if whether any crystal(s) are observed. Throughout this study, bubbles are generated in antiphase \cite{10.1121/1.3626134}, such that $\tu{B}_2$ is initiated when $\tu{B}_1$ reaches its maximum size, $X_{\tu{1,max}}$. Due to the high optical absorption coefficient of the solution ($>1000\,\mathrm{cm}^{-1}$), the incident laser energy is assumed to be fully absorbed. Further details of the optical setup, solution preparation, and experimental procedures are provided in the Methods section.

\begin{figure*}[!t]
    \centering
    \includegraphics[width=\linewidth]{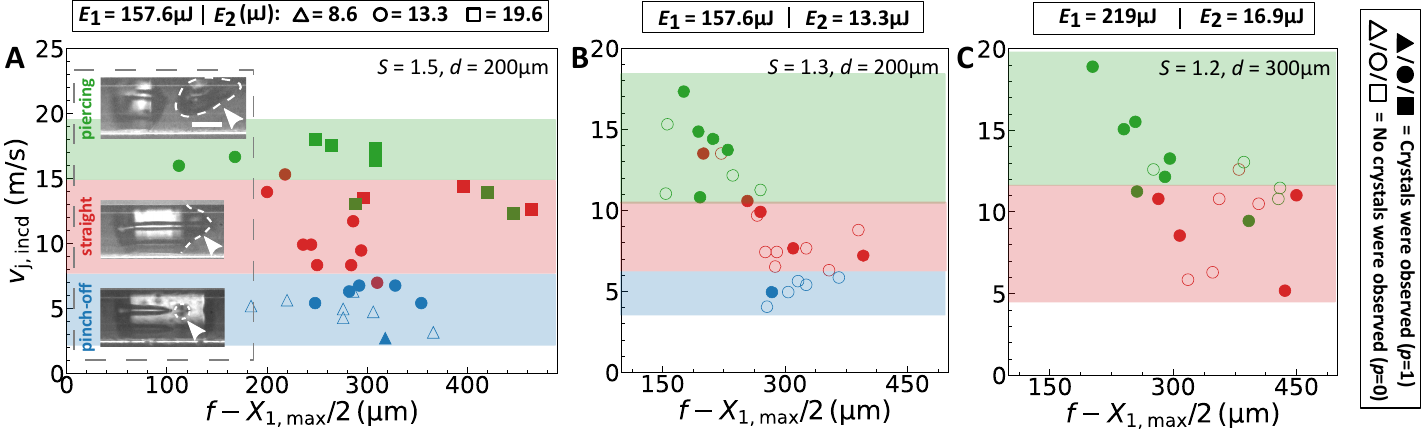}
    \caption{Incident jet velocities, jet morphology and cumulative nucleation probability. ($A-C$) Incident jet velocities ($v_\tu{j,incd}$) inside $\tu{B}_1$ at different standoff distances ($f-X_\tu{1,max}/2$). The color of the markers represents the jet morphologies with distinct shapes, breakup behavior, and temporal evolution: pinch-off, straight, and piercing, as annotated using the respective colors. In addition, each jet morphology region is shaded using the respective colors with illustrative images of $\tu{B}_1$. The shades serve as a guide to the eye. The filled markers represent the experiments where crystals were observed inside the capillary after a fixed observation time, $\tu{t}=2$\,min from the laser irradiation. Correspondingly, the empty markers represent experiments with no crystals observed at $\tu{t}=2$\,min. Whether or not crystals were observed at this fixed observation time in an experiment is used to assign a binary value for probability, $p=1\,\tu{or}\,0$ for trials with and without crystal(s), respectively. The scalebar over $\tu{B}_1$ image in panel $A$ is 100\,$\upmu$m.}
    \label{Figure_3}
\end{figure*}

\Cref{Figure_1}$B$ shows representative images from experiments, demonstrating the formation of a jet along with subsequent nucleation and growth of crystals from the supersaturated solution. The high-speed camera operating at 112,500 frames per second (fps) captures the bubble and jet dynamics, while the low-speed camera recording at 3\,fps captures the crystal growth over seconds. The experiments are carried out on an aqueous \ch{KMnO4} solution with supersaturation $S=1.5$ relative to the solubility of 0.076\,g/g of water at 298\,K \cite{CRC}. \Cref{Figure_1}$C$ highlights the formation of $\tu{B}_2$ at $t_{\tu B_2} = 240\,\upmu$s as $\tu{B}_1$ reaches its maximum size. Consequently, as $\tu{B}_2$ continues to grow, we notice the formation of a jet within B$_1$ along the axis connecting the bubble centers. This fast-moving jet pierces the right side end of $\tu{B}_1$, creating significant shear upon impact, and even facilitating the emergence of cavity and daughter bubbles. After the collapse of $\tu{B}_1$ and $\tu{B}_2$, the residual daughter bubble(s) coalesce or dissolve over a millisecond to second time range. Crystals of observable size then emerge surrounding the residual bubble(s) over seconds. The formation of crystals was found to be independent of the daughter bubbles, discussed later in \Cref{Figure_3}, where even ``non-piercing'' jets lead to crystallization.

No crystallization was observed when only a single thermocavitation bubble, $\tu{B}_1$, was generated under otherwise identical conditions. Moreover, increasing the laser energy delivered by a factor of two likewise failed to induce crystallization under a single thermocavitation bubble event, see \SI{1}. This result rules out laser-induced nucleation facilitated by local evaporation around a thermocavitation bubble as a possible mechanism - a phenomenon previously shown to promote crystal nucleation \cite{10.1103/PhysRevLett.131.124001}. The absence of nucleation with a single thermocavitation bubble points to the microjet that arises from bubble-bubble interactions as the key ingredient governing crystallization. Therefore, we investigate how the velocity of the jet, $v_\tu{j}$, influences the onset of nucleation. Since the jet velocity governs both the probability and the kinetics of crystal nucleation, as discussed later, establishing a quantitative relationship between the jet velocity and the experimental parameters provided in \Cref{Figure_1} is essential. Such a relationship would enable the rational design of laser excitation and confinement conditions to achieve desired nucleation outcomes. We employ a numerical model to resolve the full jet dynamics and determine the jet impingement velocity on the right side (opposite interface) of $\tu{B}_1$, which constitutes the key hydrodynamic parameter controlling crystallization.\\ 
% Although the analytical model of Peters \etal \cite{10.1017/jfm.2013.26} predicts the maximum velocity of the jet, it does not describe the temporal evolution of the jet

\begin{figure*}[!t]
	\centering
	\includegraphics[width=\linewidth]{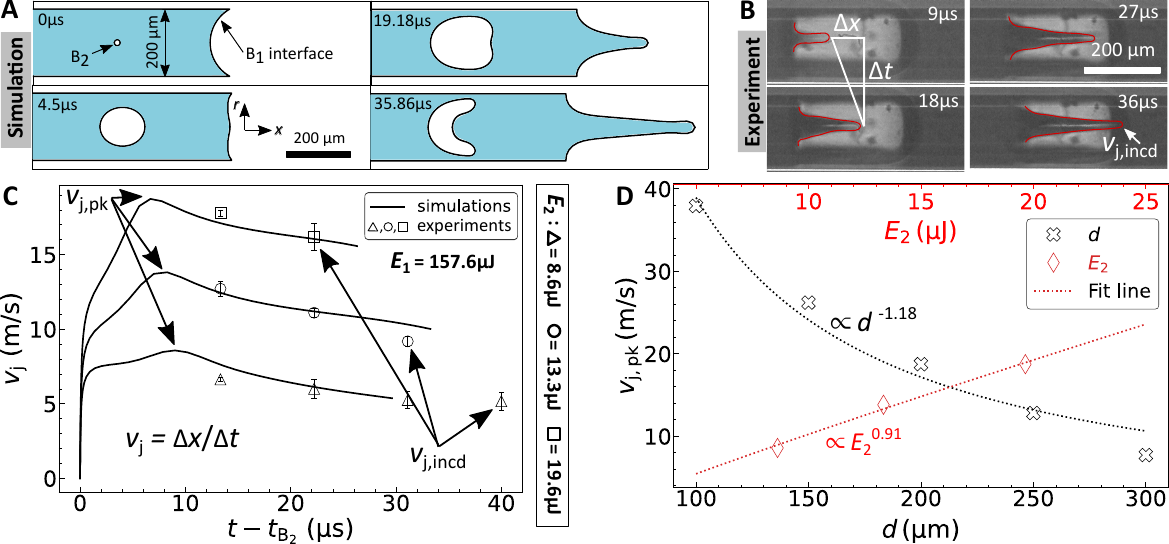}
	\caption{Temporal jet evolution and dependence of jet velocity on experimental parameters. ($A,B$) An illustration from Boundary Integral Element simulation and high-speed microscopy experiment showing jet displacement over time in a $200\,\upmu$m capillary and $E_2=19.6\,\upmu$J. ($C$) Measured temporal jet tip velocity at a fixed stand-off distance, $f-X_\tu{1,max}/2\approx 275\,\upmu$m, for three different $\tu{B}_2$ laser energies, $E_2$, and for $\tu{B}_1$ laser energy, $E_1 = 157.6\,\upmu$J. The discrete points are measurements from experiments and the continuous curves are from simulations. The incident jet velocity, $v_\tu{j,incd}$, is defined as the velocity at which the jet reaches the right side end of $\tu{B}_1$ (also illustrated in panel $B$). The peak jet velocity, $v_\tu{j,pk}$, is the maximum velocity a jet can achieve during its evolution. ($D$) Peak jet velocity ($v_\tu{j,pk}$) as a function of capillary diameter ($d$) and absorbed laser energy for $\tu{B}_2$ ($E_2$) from simulations. The three data points for $E_2$ correspond to the examples from panel $C$.}
	\label{Figure_2}
\end{figure*}
%The average standard error over mean for $E_2$ is $\pm 0.3\,\upmu$J
\noindent \textbf{Numerical simulation}\\ 
\indent We solve for the liquid momentum using an axisymmetric Boundary Integral Element Method, allowing us to simulate the interactions between two bubbles perfectly centered along the capillary axis. The rate of change in the position of the $\tu{B}_1$ interface is used to estimate the temporal evolution of the jet velocity. To decrease the code runtime, we reduce the computational domain by simulating the bubble-bubble interaction only using the near side of $\tu{B}_1$, see \Cref{Figure_2}$A$. This resembles $\tu{B}_2$ interacting with a meniscus, with the option to define the initial curvature of the $\tu{B}_1$ interface using a contact angle. In the literature, there is a numerical model for the bubble-meniscus interaction in a capillary partially filled with liquid \cite{10.1017/jfm.2013.26}, but that model fails to predict the influence of changes in capillary diameter. Moreover, bubble initialization in the model from the literature \cite{10.1017/jfm.2013.26} employs a pressure pulse that cannot be directly correlated to experiments in this study. We address these shortcomings by introducing an initial bubble of certain pressure and volume proportional to the laser energy. Using simulations we explain the observed jet formation and measured jet velocities.

In studies involving bubble-meniscus interaction in a capillary, the contact angle of the meniscus with the wall is observed to influence the jet velocity due to flow focusing. The velocity is reported to decrease with increasing contact angle \cite{10.1017/jfm.2013.26,10.1103/PhysRevX.2.031002}. Thus, the initiation of the $\tu{B}_1$ interface requires an estimate of its contact angle with the capillary wall. Using the angles measured from the images in our experiments, we impose an initial contact angle of $30^{\circ}$, consistent with others using borosilicate glass \cite{10.1364/OME.3.000284,10.1111/ijag.16571}. We assume that the gas inside the bubble consists of non-condensable gas and water vapor. The partial pressure due to the non-condensable gas is assumed to undergo a polytropic process with $\lambda=1.25$ as the polytropic index for underwater exploding bubbles \cite{Cole}. Thus, the gas pressure over the liquid at the interface can be written as \cite{10.1017/S0022112092000387},
\begin{equation}
    p_\tu{l} = p_\tu{c}  + p_0(V_0/V_\tu{b})^\lambda - \sigma_\tu{gl} \kappa.
\end{equation}
where $p_\tu{c}$ is the saturated vapor pressure of the condensable gas, $V_\tu{b}$ is the instantaneous bubble volume, $p_0$ and $V_0$ are the initial non-condensable gas pressure and bubble volume, respectively, and $\sigma_\tu{gl}$ is the gas-liquid surface tension. $\kappa = 0.5(\kappa_1 + \kappa_2)$ is the mean local curvature computed using the following equations for an axisymmetric surface \cite{10.1007/s00170-011-3165-1}: $\kappa_1 = r' x''-x' r''\,\,\tu{and}\,\, \kappa_2 = x'/r$. The superscripts $'$ and $''$ denote the first and second derivatives with respect to $\mathit{\Gamma}$ - the arc length of the piecewise smooth interface comprising the bubble surface, free surface and rigid boundary.

We treat the liquid as inviscid and incompressible and the flow induced by the bubble motion as irrotational. Using the theory for potential flow, the potential $\phi$ is related to the velocity vector $\bm{u}$ as $\bm{u} = \grad\phi$, which itself satisfies the Laplace equation,
\begin{equation}
    \laplacian \phi = 0,
\end{equation}
in the flow domain $\mathit{\Omega}$. For a sufficiently smooth function $\phi$ defined over the piecewise smooth surface $\mathit{\Gamma}$, Green's integral formula for a three-dimensional space can be written as \cite{taib1985},
\begin{multline}
c(p) \phi(p) = \int_\mathit{\Gamma} \pdvn{}\hspace{-0.3em}\left(\phi(q)\right) \, \frac{1}{|p-q|} \tu{d}\mathit{\Gamma} \\
- \int_\mathit{\Gamma} \phi(q)\, \pdvn{}\hspace{-0.3em}\left(\frac{1}{|p-q|}\right) \tu{d}\mathit{\Gamma},
\label{biemeq}
\end{multline}
where $p$ is a field point anywhere in $\mathit{\Omega+\Gamma}$, $q \in \mathit{\Gamma}$, $n$ is the outward unit normal, and $c(p)$ equals $4\pi$ if $p \in \mathit{\Omega}$ or $2\pi$ if $p \in \mathit{\Gamma}$. Choosing $p$ on $\mathit{\Gamma}$ results in a solution for either $\phi$ or $\partial \phi/\partial n$ on $\mathit{\Gamma}$ if the other is specified. Following a Lagrangian framework, the kinematic and dynamic boundary conditions on the bubble surface are written as
\begin{gather}
    \Delta r = (\partial \phi/\partial r)\, \Delta t, \hspace{1em} \Delta x = (\partial \phi/\partial x)\, \Delta t, \hspace{1em} \tu{and}\\
    \Delta \phi = \left[\frac{1}{2}|\grad \phi|^2 + \frac{p_\infty-p_l}{\rho_\tu{L}}\right] \, \Delta t,
\end{gather}
where $p_\infty$ is the pressure as $x\to\infty$ and $\rho_L$ is the liquid density.

With $\Delta p = p_\infty-p_c$, we use the following characteristic scales to non-dimensionalize the associated parameters in simulation: $R_\tu{max}(\rho_\tu{L}/\Delta p)^{1/2}$ for time, $R_\tu{max}(\Delta p/\rho_\tu{L})^{1/2}$ for potential and $(\Delta p/\rho_\tu{L})^{1/2}$ for velocity. This implies that we non-dimensionalize lengths using $R_\tu{max}$, which is the maximum radius of the spherical bubble when simulated in an unbounded liquid domain - with no capillary walls or $\tu{B}_1$ surface. Correspondingly, the change in potential on the B$_1$ and B$_2$ surface are given by \cite{10.1016/j.enganabound.2012.04.009,10.1007/s10409-012-0147-y},
\begin{gather}
    \Delta \widehat{\phi} = \left[1 + \frac{1}{2} |\bm{\widehat{u}}|^2 - \zeta (V_0/V_\tu{b})^\lambda + \alpha\widehat{\kappa} \right]  \,\Delta \widehat{t} \hspace{1em}\tu{and}\\
    \Delta \widehat{\phi} = \left[\frac{1}{2} |\bm{\widehat{u}}|^2 + \alpha\widehat{\kappa} \right]  \,\Delta \widehat{t},
\end{gather}
respectively. Here $\zeta=p_0/\Delta p$ is the strength parameter and $\alpha = \sigma_\tu{gl}/(R_\tu{max}\Delta p)$ is the surface tension parameter. The superscript $\,\widehat{}\,$ denotes the dimensionless form of the parameters.

The bubble initialization requiring $p_0$ and $V_0$ is theoretically related to the initial potential energy of the bubble with $p_0V_0 \propto R_\tu{max}^3$ \cite{10.1017/S0022112098001207}. \SI{2} discusses the validation of the initial conditions together with the analytical equation for the estimation of $\zeta$. It is essential to note that only two of the three dimensionless parameters $\zeta$, $R_0 /R_\tu{max}$ and $\lambda$ can be chosen independently. In all our analysis, upon fixing the non-dimensionalized initial radius $R_0/R_\tu{max}$ as $0.1$, we get $\zeta=313$. The chosen value for $\zeta$ is within the $100-500$ range for exploding bubbles \cite{Cole}, with the change in value within this range having negligible effect on bubble/jet dynamics, similar to others \cite{10.1017/S002211200700852X}. The numerical solution procedure to \Cref{biemeq} can be found in \SI{2}.

\Cref{Figure_2}$A-B$ shows the representative images for jet evolution at different time instances from experiments and simulations. The velocity $v_\tu{j}$ of the jet tip is determined based on the rate of change in the position of the jet tip. \Cref{Figure_2}$C$ provides the temporal jet tip velocity for a chosen standoff distance, $f-X_\tu{1,max}/2\approx 275\,\upmu$m, for three different laser energies $E_2$ of $\tu{B}_2$. In all three cases, we first see the jet accelerate due to the momentum gain from $\tu{B}_2$ formation and bubble-bubble interaction. After reaching a maximum velocity, $v_\tu{j,pk}$ - peak jet velocity, the jet then decelerates due to surface tension counteracting the increase in surface area caused by the evolution of jet \cite{10.1017/jfm.2013.26}. The incident velocity, $v_\tu{j,incd}$ is the velocity at which the jet strikes the right side of the first bubble, $\tu{B}_1$. It should be noted that $v_\tu{j,pk}$ is a characteristic velocity of the jet, while for a fixed $v_\tu{j,pk}$, $v_\tu{j,incd}$ depends on $X_\tu{1,max}$ (or $E_1$ and $d$). Thus, for an evolving jet $v_\tu{j,incd}$ increases with decreasing $X_\tu{1,max}$, making $v_\tu{j,pk}$ the right parameter to study jet characteristics in line with the literature \cite{10.1017/jfm.2013.26}. However, $v_\tu{j,incd}$ is the necessary parameter to study crystallization characteristics, discussed later in the following section.\\  

\noindent \textbf{\large{Discussion}}\\
\indent In \Cref{Figure_2}$C$, the simulations show quantitatively similar jet evolution behavior when compared with high-speed microscopy experiments. However, there are some differences. The lifetime of bubble $\tu{B}_2$ in the simulation from \Cref{Figure_2}$A$ ($\approx 36\,\upmu$s) is shorter than in the experiments in \Cref{Figure_2}$B$ [$O (100\,\upmu$s)]. We attribute the difference to neglection of viscous effects and capillary forces between the interface of $\tu{B}_1$ and the wall. Yet, as seen in \Cref{Figure_2}$D$, the model accurately predicts that $v_\tu{j,pk}$ has a power-law dependence on $d$ and $E_2$ with exponents of $-1.18$ and $0.91$, respectively, closely consistent with previous observations \cite{10.1103/PhysRevX.2.031002}. Moreover, results presented in \Cref{Figure_2}$D$ are in line with physical intuition. As $d$ decreases with other experimental parameters held constant, the peak jet velocity increases as capillary walls confine and direct momentum along the tube. Likewise, increasing $E_2$ provides more momentum for the jet, resulting in higher jet velocities. Furthermore, the $R_\tu{max}$ from simulations directly correlates to the laser energy from experiments. The determined power law fit of $E_2 \propto R_\tu{max}^{2.603}$ (see \SI{2}) is consistent with the reported empirical relation for energy difference above a threshold laser energy \cite{10.1103/PhysRevLett.88.078103}. The numerical model discussed thus quantitatively captures the bubble-bubble interaction.

An early observation from high-speed microscopy recordings is the distinct jet morphologies shown as insets in \Cref{Figure_3}$A$. Based on the jet velocity, the jets show different morphologies with distinct temporal evolutions and breakup behavior. Therefore,  the observed jet morphologies in experiments are categorized into three types: (i) pinch-off: jet breaks into droplet(s); (ii) straight: jet does not break up before it reaches the opposite end of $\tu{B}_1$, and therefore will deform the incident vapor-liquid interface but will not pierce it; and (iii) piercing: the jet pierces the opposite end of $\tu{B}_1$, leading to the creation of daughter bubble(s), see \Cref{Figure_3}$A$. The tendency of a jet to break up is evaluated using the Weber number $\textit{We} = \rho_\tu{L} v_\tu{j}^2 r_\tu{j}/\sigma_\tu{gl}$, with $r_\tu{j}$ the jet radius. Jetting is reported to occur for $\textit{We}>4$ \cite{10.1063/1.3524533,10.1146/annurev-matsci-070909-104502}. In this work, \textit{We} ranges between $4-65$, satisfying the necessary condition for jet formation and breakup. However, as the distance the jets can travel within $\tu{B}_1$ is limited to $X_\tu{1,max}$, jets with relatively higher velocities possess sufficient momentum to reach the right side end of $\tu{B}_1$ before breaking up, while the jets with lower velocities break up (pinch-off) due to the Rayleigh instability \cite{10.1063/1.3524533}. The observed velocity range ($4-8\,$m/s) for jet pinch-off is in agreement with the reported satellite-free drop generation range \cite{10.1103/PhysRevApplied.6.024003}. A clear observation from \Cref{Figure_3}$A$ is that $v_\tu{j}$ increases with increasing laser energy, $E_2$, regardless of jet morphology at fixed $f-X_\tu{1,max}/2$, $E_1$ and $d$. This experimental observation is in line with physical intuition. The higher $E_2$, the more laser energy is available for momentum, resulting in higher jet velocities. The corresponding number of crystals observed for the experiments presented in \Cref{Figure_3} is reported in \SI{3}.
%Arguably, the jet could increase the solute transport due to the transfer of momentum, favoring crystal nucleation \cite{10.1021/acs.cgd.5b01437}. However, this is a mere hypothesis with limited experimental evidence in literature hence we attempt to further decipher how the emergence of the jet might alter the crystal nucleation phenomena.
 
\Cref{Figure_3} shows the incident jet velocities for a range of supersaturations $S=\{1.5,1.3,1.2\}$, laser energies $E_1$ and $E_2$, and capillary diameters $d=\{200,200,300\}\,\upmu$m, respectively, for varying standoff distances. At the same stand-off distance we observe variation in the jet velocity between experiments in \Cref{Figure_3}. We attribute this variation to the experimental difficultly in maintaining the left side end curvature (contact angle) of $\tu{B}_1$ due to contact angle pinning and instabilities at the free interface. As \Cref{Figure_3} classifies different jet morphologies, it can shed light on the question ``Does jet morphology influence nucleation probability and kinetics?''. To this end, we introduce the cumulative nucleation probability at a fixed time lag, $p$, as the measure of the likelihood of crystal formation due to microjet formation. Throughout this work, whether a solution did crystallize ($p=1$, closed symbols)  or not ($p=0$, open symbols) was determined at a fixed observation time of 2\,minutes. The results presented in \Cref{Figure_3} show a clear dependence of $p$ on $v_\tu{j,incd}$ (incident jet velocity), $S$ (solution supersaturation) and $d$ (capillary diameter), with no influence of jet morphology. Apparent lack of a direct influence of jet morphology on nucleation probability is noteworthy, as one might intuitively expect morphologies with larger interfacial surface areas to promote heterogeneous nucleation and thereby enhance crystallization kinetics. At first glance, \Cref{Figure_3}$A$ appears to indicate a morphology dependence, with most non-nucleating experiments with $p=0$ clustered within the pinch-off regime (with blue shade). However, closer inspection of \Cref{Figure_3}$B,C$ shows that this notion is misleading, as open and closed symbols are scattered across different morphologies with no distinct preference. Moreover, pinch-off type jets possess the highest surface-area-to-volume ratio with a pinched-off droplet. Therefore, this morphology should exhibit the highest nucleation probability if heterogeneous nucleation at the vapor-liquid interfacial area were the dominant mechanism. However, the opposite trend is observed. Instead, nucleation correlates primarily with incident jet velocity, suggesting that jet-induced shear, rather than jet morphology, governs crystallization.  

\begin{figure*}[!t]
    \centering
    \includegraphics[width=\linewidth]{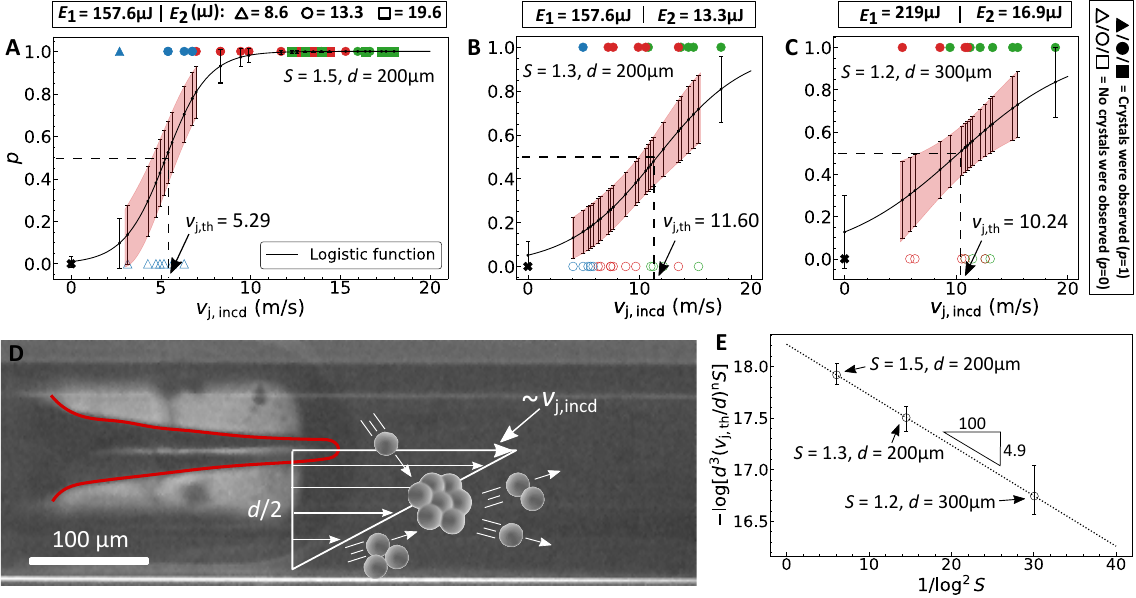}
    \caption{Cumulative nucleation probability and calculation of solute-solution interfacial tension. ($A-C$) The logistic regression function plot for the panels $A-C$ in \Cref{Figure_3}, respectively. $v_\tu{j,th}$ is the threshold velocity for crystal formation above which the cumulative nucleation probability, $p\geq 0.5$. The errorbars are the standard errors. The representation of the markers is similar to \Cref{Figure_3}. ($D$) Schematic of the cluster formation via particle addition and removal processes in the presence of shear flow. The arrows represent the linear velocity streamlines. ($E$) The plot shows the estimates from \Cref{gamma}. Appropriate values for the variables are supplied for the parameters discussed in this work, $S=\{1.5,1.3,1.2\}$, $d=\{200,200,300\}\,\upmu$m and $v_\tu{j,th}$ from panels $A-C$, respectively. The slope corresponds to the value of $B$ in \Cref{gamma}, with which we estimate the solute-solution interfacial tension $\sigma_\tu{sl}$. The error associated with the $v_\tu{j,th}$ is estimated from the shaded region, which corresponds to the standard errors of the logistic function in panels $A-C$.}
    \label{Figure_4}
\end{figure*}

We see an increase in the experiments resulting in crystallization ($p=1$, closed symbols) with increasing $v_\tu{j,incd}$, $S$ and $d$ across all panels in \Cref{Figure_3}. Therefore, a characteristic parameter to compare experiments conducted under different experimental conditions is needed. In \Cref{Figure_3}, this characteristic parameter is established using a threshold jet velocity for observed crystal formation, $v_\tu{j,th}$, defined as the value above which the cumulative nucleation probability is greater than $0.5$. For each set of $S$ and $d$ employed in this work, the $p$ as a function of $v_\tu{j,incd}$ is calculated using the logistic regression as shown in \Cref{Figure_4}$A-C$. We report an increase in $v_\tu{j,th}$ with decreasing supersaturation - a larger characteristic jet velocity (or shear) is needed to trigger crystal nucleation at lower supersaturations.  Furthermore, upon increasing $d$ from 200 to 300\,$\upmu$m while maintaining $S=1.3$, we observed a steep rise in the number of crystals formed ($>20$, refer \SI{4}). Therefore, the experiments in \Cref{Figure_3}$C$ in $300\,\upmu$m capillary were performed with relatively lower supersaturation ($S=1.2$) to facilitate easy counting of the number of crystals.

The impingement of a jet onto the liquid surrounding the bubble can create shear \cite{10.1073/pnas.1518679112}, thereby altering the nucleation kinetics \cite{10.1103/PhysRevE.87.062307,10.1103/PhysRevE.93.042803}, as illustrated in \Cref{Figure_4}$D$. However, the exact mechanism of how shear facilitates nucleation is still under debate, shrouded in mechanisms involving enhanced mass transfer due to flow, agitation-enhanced cluster aggregation and shear-induced molecular alignment of clusters \cite{10.1021/acs.cgd.3c01198,10.1063/1.2981052,10.1021/acs.cgd.3c00232}. From classical nucleation theory, the nucleation rate, $J$, which is the number of nuclei formed per unit volume per unit time, is expressed as \cite{10.1002/crat.200310070},
\begin{equation}
    J = A S\exp[-B / \log^2S],
    \label{J eqn}
\end{equation}
where $B$ is a dimensionless thermodynamic parameter given by $B=16\pi v_0^2 \sigma_\tu{sl}^3/(3k_\tu{B}^3T^3)$. Here $v_0$ is the molecular volume of the solute, $\sigma_\tu{sl}$ is the solute-solution interfacial tension, $k_\tu{B}$ is Boltzmann’s constant and $T$ is the absolute temperature. $A$ is the supersaturation-independent kinetic parameter for homogeneous nucleation. The influence of liquid shear rate $\dot{\gamma}$ on the nucleation rate is accounted for in $A$ with empirical evidence showing a power law dependence, $A\propto\dot{\gamma}^n$ \cite{10.1021/cg5008878,10.1021/acs.cgd.9b01584,10.1021/cg4007636,10.1016/j.polymer.2016.02.059,10.1039/D3SM00528C}. In addition, for aqueous glycine, Forsyth et al. \cite{10.1021/acs.cgd.5b01042} demonstrated $A$ to positively scale with the solution container's surface area $\Pi$ as $A\propto \dot{\gamma}\Pi$. Combining the relations, we get the empirical correlation $A \propto (v_\tu{j}/d)^n d$, with $\dot{\gamma} \sim v_\tu{j}/d$ (see \Cref{Figure_4}$D$) and $\Pi \sim \pi d$ over unit length of the capillary. This thus establishes the relationship to the governing parameters accounted for in the pre-exponential kinetic factor $A$, while keeping the thermodynamic parameters $B$ and $S$ constant.

%Similar to shearing flow, the possibility of an elongational flow (due to the motion of the jet along the $x$-axis) to alter the crystal nucleation kinetics cannot be ignored \cite{10.1021/acs.macromol.3c02632}. Therefore a scaling $v_\tu{j}/X_\tu{1,max}$ similar to shear rate is improvised to analyze its magnitude. The value for this adapted scaling is at least 1.7 times less than $\dot{\gamma}$ - the shear rate, since $X_\tu{1,max}/d>1.7$ in this work. We therefore ignore the effect of elongational flow on nucleation rate in this work.

By definition, $J$ can be expressed as $N/(t_\tu{obs}V)$, where $N$ is crystal count, $t_\tu{obs}$ is observation time and $V$ the capillary volume. We establish the crystallization threshold criterion as $N\approx 1$, with a fixed observation time of $t_\tu{obs}=2\,\tu{min}$. Therefore, near the crystallization threshold, this makes $J$ only a function of capillary volume, $J\propto1/(\pi d^2 L)$. Since we maintain a constant capillary length ($L$) throughout this work, we simplify, 
\begin{equation}
    J/A = S\exp[-B / \log^2S] \propto d^{-3} (v_\tu{j,th}/d)^{-n} / C.
    \label{J/A}
\end{equation}
$C$ is a constant whose value is based on $N$, $t_\tu{obs}$, $L$, temperature, and solute-solvent properties \cite{10.1002/crat.200310070}. The thermal penetration depth in the liquid surrounding the bubble is calculated using the lifetime of the bubbles, which is $O(10\,\upmu\tu{m})$. However, the momentum penetration depth due to jet impingement, calculated using Newton's approximation of impact depth \cite{marsden2017}, is $O(100\,\upmu\tu{m})$. We therefore neglect the effect of local temperature variations in the liquid surrounding the bubble in \Cref{J/A} and take $T=298\,\tu{K}$ throughout our analysis. Correspondingly, we neglect the change in $S$ due to $T$. Moreover, the solute penetration depth that would exist due to exchange of solvent at the $\tu{B}_1$ ends (gas-liquid interface) \cite{10.1103/PhysRevLett.131.124001} is calculated to be $O(1\,\upmu\tu{m})$. This, however, is much less than the momentum penetration depth due to microjets, and therefore allows us to ignore the solute boundary layer surrounding $\tu{B}_1$ in our calculations. As a result, we take $S$ as the supersaturation in the bulk. By rearranging \Cref{J/A}, we get
\begin{equation}
    -\log[d^3(v_\tu{j,th}/d)^n S] = -B / \log^2S + \log C.
    \label{gamma}
\end{equation}

With three unknowns: $n$, $B$, and $C$, in the above equation, at least three datasets of the other variables are required to compute the unknowns. To this end, we condense all the data presented in \Cref{Figure_4}$A-C$ into three datasets. Thus for $S=\{1.5,1.3,1.2\}$ and $d=\{200,200,300\}\,\upmu$m, the corresponding $v_\tu{j,th}$ discussed in \Cref{Figure_4}$A-C$ are respectively used. This results in $n=0.71$ and $B=0.049$. The estimates after substituting the appropriate values of variables in \Cref{gamma} are shown in \Cref{Figure_4}$E$, in which $B$ is the slope of the fit line. This estimated value for $n$ is in close agreement with the empirically observed $1-3$ range from literature that employ various solutes and shearing techniques \cite{10.1021/cg5008878,10.1021/acs.cgd.9b01584,10.1021/cg4007636,10.1016/j.polymer.2016.02.059,10.1039/D3SM00528C,10.1021/acs.cgd.5b01042}, see \SI{5}. Furthermore, using the estimated value for $B$ from the fit, we can calculate $\sigma_\tu{sl}=2.8$\,mJ/m$^2$ at 298\,K, which is in close agreement with 3.16\,mJ/m$^2$ calculated using the Mersmann relation \cite{10.1016/0022-0248(90)90850-K}. See \SI{6} for the calculations of $\sigma_\tu{sl}$. The fact that the extracted values of $n$ and $B$ are both in agreement with literature values thus strengthens the validity of the proposed scaling of the kinetic parameter with microjet-induced shear rate in \Cref{gamma} in the context of classical nucleation theory \cite{10.1016/B978-0-444-63303-3.00032-8,Kashchiev2000}. The proposed scaling suggests that a threshold jet velocity and, by extension, a critical shear rate above which nucleation is more likely. Moreover, the threshold jet velocity above which nucleation is more likely depends also on the solution supersaturation and capillary diameter. This dependence can be rationalized within the classical nucleation theory framework, as increasing $S$ and $d$ would increase the amount of solute available to form nuclei within the volume of interest.\\  

The model solute studied, KMnO4 has an increasing solubility with temperature and therefore is expected not to crystallize due to the local heating of solvent by laser. This study thus distinguishes itself from the light-driven crystal nucleation of retrograde (inverse) soluble compounds via heating \cite{10.1038/s41467-023-42126-4}. Moreover, a single cavitation bubble triggered under identical conditions did not induce crystallization. Therefore, previously proposed solute accumulation at the vapor-liquid interface due to thermocavitation\cite{10.1103/PhysRevLett.131.124001}  cannot explain observations in this study. Thus, the observed correlation of higher cumulative nucleation probabilities with increasing jet velocity in \Cref{Figure_4}$A-C$ along with the CNT analysis presented in \Cref{Figure_4}$E$ supports the hypothesis of shear-induced nucleation emerging from hydrodynamic bubble-bubble interactions. This presents a novel route to nucleation from solution. However, a critical look into these results, as well as a careful comparison to the literature, is needed.
% These results collectively suggest that shear created by a jet emerging from hydrodynamic bubble-bubble interactions increases the nucleation rate relative to spontaneous nucleation.

The first question we should ask is:  Can shear influence the nucleation in this work? Shear induced nucleation is reported for colloids, proteins, polymers and other soft matter systems, as well as for molecular solutes \cite{10.1103/PhysRevLett.93.068303,10.1021/acs.chemrev.7b00500,10.3139/217.930236,10.1021/cg4007636}. Unlike colloids, imaging of individual molecules in solution under shear is experimentally challenging. Therefore the precise shear-induced nucleation mechanism for molecular solutes has so far remained elusive. In this work, we estimate shear rates $O(10^5\,\tu{s}^{-1})$ - among the largest reported in literature \cite{10.1021/acs.cgd.3c00232},  applied over the lifetime of the microjet $O(10^{-6}\,\tu{s})$ - the shortest reported in literature (see \SI{5}). To answer the question above, we employ the P\'eclet number, $Pe=\dot{\gamma}l^2/D$, as the critical threshold where flow-driven transport outpaces quiescent relaxation due to thermal fluctuations. $l$ is the characteristic length scale and $D$ is the diffusion coefficient. For \ch{KMnO4}, $D$ is taken as $1\times 10^{-9}$\,m$^2$/s. For the calculation of $l$, we use the critical cluster diameter of \ch{KMnO4} from CNT in the absence of any external field for the lowest supersaturation employed in this work, $S=1.2$. The critical cluster diameter is calculated as 1.64\,nm, see \SI{7} for the calculation \cite{10.1002/crat.200310070}. Using the above values, the estimated $Pe \approx O(10^{-4})$, which is significantly lower than unity - not sufficient to argue that shear is a prominent mechanism in the enhanced nucleation observed in this work. Alternatively, we can calculate to what length scale the pre-nucleation clusters need to grow so that $Pe \approx 1$. This calculation suggests a characteristic cluster size in the range of $100$ to $300$\,nm. From a thermodynamic point of view, it is problematic to have such large clusters in (metastable) equilibrium. Therefore, it is more natural to assume that the effect of shear impacts the crystal growth stage rather than the nucleation stage. In the crystal growth stage, the length scales are obviously much larger than the relevant scales in the nucleation stage.

On the theoretical side, Mura and Zaccone \cite{10.1103/PhysRevE.93.042803} developed a shear-modified classical nucleation theory (CNT) framework where nucleation is governed by two competing effects: enhanced molecular transport and nucleus straining. This framework predicted an optimal shear rate at which the nucleation rate is enhanced to an order of magnitude compared to the case in the absence of flow. Debuyssch\`ere \etal \cite{10.1021/acs.cgd.3c00232} explored the proposed framework experimentally with glycine reaching shear rates as high as $O(10^4\,\tu{s}^{-1})$ in a millimeter-sized capillary. These authors observed a maximum enhanced nucleation at shear rates  $O(10^3\,\tu{s}^{-1})$ with nucleation rate decreasing at higher shear rates as qualitatively predicted by Mura and Zaccone \cite{10.1103/PhysRevE.93.042803}. Yet, in line with our conclusion, Debuysschere \etal \cite{10.1021/acs.cgd.3c00232} also concluded that their experiments cannot be described by shear-modified CNT. 

To explain the observed shear-induced nucleation enhancement in their experiments, Debuysschere \etal\cite{10.1021/acs.cgd.3c00232} and other groups refer to the existence of non-classic prenucleation clusters that are larger than what CNT predicts \cite{10.1021/bk-2020-1358.ch002,10.1021/acs.cgd.6b00794}. As prenucleation clusters are presumed to be both larger and diffuse more slowly, $Pe$ numbers of order one can be reached, emphasizing the role of flow over diffusion. Also, due to growing indirect experimental evidence supporting the existence of non-classical clusters as large as 100 nm \cite{10.1021/acs.cgd.5b01042} for a broad range of inorganic aqueous solutions \cite{10.1002/anie.201610275,10.1021/cg300150u,10.1107/S2053230X15008997}, we acknowledge our results could be interpreted in this context. However, the exact physicochemical composition of these clusters is experimentally challenging to pinpoint. We also acknowledge that such large prenuclueation clusters ar problematic conceptually. For our system, this would mean equilibrium cluster sizes involving $10^8 - 10^9$ ions, which is theoretically very difficult to accept without having to invoke some unknown stabilization mechanism. Therefore, we would like to also propose alternative explanations for our observations. 

One factor that may play a role in the enhanced nucleation is the supersonic airflow due to the cavity collapse (see \Cref{Figure_1}$B$ and \SI{8}). The supersonic airflow generated at the impacted $B_1$ interface would enhance the $\dot{\gamma}$ in $Pe$ \cite{10.1103/PhysRevLett.96.154505,10.1103/PhysRevLett.102.034502}. The supersonic airflow exiting the cavity formed upon microjet impact closely resembles that of compressible flow through a nozzle - where here the ``nozzle'' is a liquid cavity shrinking rapidly in time \cite{10.1103/PhysRevLett.104.024501}, see \SI{8}. While the presence of the cavity is evident for piercing jets, the straight jets however due to its relatively low velocities would create cavities with extremely short length and time scale - which are experimentally challenging to capture. Overall, this supersonic airflow, extensively studied in literature \cite{10.1103/PhysRevFluids.3.110504,10.1063/1.1564347}, with velocities $O(1000-10000\,\tu{m/s})$ over the liquid nozzle can create larger shear rates [$O(10^8\,\tu{s}^{-1})$] (see \SI{8} for calculation). These transient shear rates are three orders of magnitude larger than what we predict on the basis of the incident jet velocity, and may create localized hydrodynamic environments in which advective transport and mechanical straining of much smaller pre-nucleation clusters become significant \cite{10.1103/PhysRevE.93.042803}. At the same time, the timeframe in which the shear rate is so high is also very small. Whether it is long enough to allow for formation of embryonic nuclei is unclear, but the potential for enhanced nucleation is obviously there during the cavity collapse.    

Another mechanism that may contribute to the enhanced nucleation rate involves the coupled concentration and temperature fields surrounding the bubble. The expanding bubble is accompanied by both a solute-rich accretion layer, approximately $O (100\,\mathrm{nm})$ thick, and a thermal boundary layer extending several micrometers into the liquid \cite{10.1103/PhysRevLett.131.124001}. Because the jet originates from liquid adjacent to the bubble interface, the strong extensional flow associated with jet formation is expected to stretch and thin both layers. Upon impact and coalescence with the opposite meniscus, the reduced thickness of the thermal boundary layer may enhance heat transfer away from the concentrated accretion zone, resulting in local cooling. Since the accretion layer already possesses an elevated solute concentration, even a modest temperature decrease could transiently increase the local supersaturation, thereby providing an additional thermodynamic driving force for nucleation alongside the hydrodynamic effects discussed above. Just like the supersonic airflow created due to the transient cavity collapse explanation, confirming thermal effects is experimentally challenging due to the time and length scales involved in microjet evolution. Yet these two explanations could work in concert with the shear-induced nucleation hypothesis.  

An observation worth mentioning is the different induction times reported in the literature compared to this study. The induction time for crystallization is defined as the time elapsed from the moment of shear until the first crystal is detected. In this work, we observe an induction time in the order of milliseconds to seconds - restricted within the fixed observation of 2\, minutes. However, the induction times reported for different solutes and shearing methods in the literature is in orders of tens of seconds to minutes as tabulated in \SI{5} \cite{10.1021/cg5008878,10.1021/acs.cgd.9b01584,10.1021/cg4007636,10.1016/j.polymer.2016.02.059,10.1039/D3SM00528C,10.1021/acs.cgd.5b01042}. It should also be noted that the maximum shear rate reported in literature is two orders of magnitude lower than this work, $O(10^5\,\tu{s}^{-1})$. While, in literature, the applied shear spans the whole induction time when compared to a very short time $O(10^{-6}\,\tu{s})$ in this work - the lifetime of the microjet. This characteristic of microjet-induced shear described in this work thus offers a unique opportunity to separate the nucleation stage from the crystal growth stage. As the controlled shear experiments in the literature apply the shear for a prolonged time $O(10\,\tu{s})$, it is very likely that the growth stage is affected by the shear. 
%In the case of jet-induced nucleation, the timescales where shear is high are very short compared to the literature. 

%In summary, we have demonstrate a novel crystallization route via microjets forming emerging from bubble-bubble hydrodynamic interaction in supersaturated aqueous solutions.We correlate the observed enhanced nucleation in experiments to the velocity of jet and the shear generated upon jet impact. With the effect of shear embodied into the kinetic prefactor in classical nucleation theory, we report a threshold jet velocity, only above which nucleation is significant. This threshold velocity increases with decreasing supersaturation and capillary diameter. The proposed approach of shearing the liquid in a controlled environment also avoids nuclei formation at interfaces due to prolonged exposure \cite{10.1021/acs.jpclett.0c00540}. Combining experiments with simulations, we engineer the jet velocity responsible from observed nucleation enhancementsuggesting an previously unreported hydrodynamic route to alter nucleation. This technique also enables primary control over crystal nucleation at lower laser energies and solution supersaturation compared to crystallization via a single bubble opening routes to manufacturing of crystalline materials previous considered difficult to manufacture in crystalline form. The proposed study sheds light on how hydrodynamic interactions between bubbles could alter observed nucleation rates in applications ranging from hydrothermal mineralization in deep-sea vents to industrial manufacturing of crystalline materials.\\

In summary, we demonstrated a crystal nucleation mechanism driven by microjets generated through bubble-bubble hydrodynamic interactions in a supersaturated aqueous solution of a model salt. We identified a threshold jet velocity above which nucleation becomes significant and showed that this threshold increases with decreasing supersaturation and capillary diameter in experiments. By combining experiments, simulations, and theory, we propose microjet-induced shear as the mechanism promoting crystal nucleation. More broadly, this work reveals bubble-bubble hydrodynamic interactions as a viable route for flow-assisted control of nucleation kinetics and phase transitions - a previously overlooked phenomenon. The prospective applications range from geothermal magma flows \cite{10.1126/science.adw8543} to next-generation manufacturing technologies for pharmaceuticals\cite{10.1038/s44286-025-00268-w} and high-performance perovskite solar cells \cite{10.1126/science.adt5001}.\\

% The proposed scaling relations for jet velocity with absorbed laser energy, bubble-bubble stand-off distance and capillary diameter using axisymmetric boundary integral element method simulations agree with the experiments performed and with literature

%  Furthermore, we numerically simulate the bubble-bubble interaction and the resulting evolution of jet using axisymmetric boundary integral element method. The concomitant numerical study is used to determine the characteristic scaling laws for jet velocity on bubble-bubble standoff distance, capillary diameter and laser energy.

\noindent\textbf{\large Methods}\\
\noindent \textbf{Optical setup}\\ 
\indent For sample illumination, we used white LED light (KL1500, Zeiss). An infinity-corrected 10x objective (Plan Fluor, Nikon) with 0.3 numerical aperture was used for laser 1. An 8x objective (Planapochromat, Carl Zeiss) with 0.1 numerical aperture was used for laser 2. The 10x objective was also used for imaging with resolution $\approx 2\,\upmu\tu{m}/\tu{px}$ for a high-speed camera (FASTCAM NOVA S16, Photron) and $\approx 0.74\,\upmu\tu{m}/\tu{px}$ for a low-speed camera (Imager Pro LX 16M, LaVision). The basis on how to build the setup with bill of materials is reported elsewhere \cite{10.1016/j.ohx.2023.e00415}. Both lasers 1 and 2 are pulsed Nd:YAG lasers with the configurations listed in \Cref{laser parameters}.
\begin{table}[!th]
  %\small
  \centering
  \caption{Laser parameters.}
    \begin{tabular}{>{\centering\arraybackslash}p{10em}>{\centering\arraybackslash}p{7em} >{\centering\arraybackslash}p{8em}} \hline
    \textbf{Parameter} & \textbf{Laser 1} & \textbf{Laser 2} \\ \hline
    Model & Nano L 90-100 & Powerlite DLS 8000, Continuum\\
    Wavelength & 532\,nm & 532\,nm\\
    Pulse duration & 5\,ns & $7-9$\,ns\\
    Diameter at source & 4\,mm & 9\,mm\\
    Diameter at objective & 8\,mm & 3\,mm\\
    TEM$_{00}$ & yes & no \\
    \hline
    \end{tabular}%
    \label{laser parameters}
\end{table}\\

\noindent \textbf{Solution preparation}\\ 
\indent Ultrapure water (18.2\,M$\Omega$\,cm, ELGA Purelab) was used to prepare the supersaturated \ch{KMnO4} (223468, Sigma-Aldrich) solutions in an 8\,ml vial (SF8, BGB). A solubility of $7.6\,$g/100g-H$_2$O at 298\,K for \ch{KMnO4} was used \cite{CRC,baxter}. After preparation, the solution was stirred at 1000\,rpm over a hot plate maintained at 333\,K for one hour to accelerate the dissolution. Then the solution was placed in the oven maintained at 323\,K overnight ($>12\,$hours) for complete dissolution. During the experiments, the vials were placed over a hot plate maintained at 333\,K and constantly stirred at 1000\,rpm.\\

\noindent \textbf{Experimental protocol}\\ 
\indent For each experiment, the solution was freshly loaded into a new capillary. The warm solution from the hot plate was transferred using a micro-pipette. The warmth prevents spontaneous nucleation while handling. Once the supersaturated solution was transferred, the capillary ends were covered using droplets of silicone oil (378321, Sigma-Aldrich) to prevent evaporation. The step-by-step experimental procedure during each trial is detailed in \Cref{experimental protocol}.
\begin{table}[!ht]
  \small
  \centering
  \caption{Experimental protocol}
    \begin{tabular}{>{\arraybackslash}p{20.5em}>{\centering\arraybackslash}p{5.5em}} \hline
    \textbf{Description of the event sequence} & \textbf{Time sequence (minutes)} \\ \hline
    1. Fill the capillary with supersaturated solution & at $0$ \\
    2. Place the filled capillary over the sample stage in the microscope and allow it to cool down. The LED light is switched off during this process. & $0-2$ \\
    3. Scan the capillary end to end to check if any crystals have formed. If there are no crystals, then align the capillary with its mid along the length over the objective. & $2-6$ \\
    4. Trigger a few laser pulses with the shutter closed to warm up the lasers. & $6-6.5$ \\
    5. Prepare the camera and oscilloscope to wait for the trigger. The laser is shot at $\approx 7\,$min. & $6.5-7$ \\
    6. The high-speed camera records for 3\,seconds, while the low-speed camera continues recording until 2\,min. After the $5^\tu{th}$ second, the intensity of the LED is kept at a minimum, as the large exposure time of the low-speed camera allows in more light. & $7-9$ \\
    %7. The capillary is scanned end to end to check for crystal(s). & $9-11$ \\
    \hline
    \end{tabular}%
    \label{experimental protocol}
\end{table}\\

\noindent \textbf{\large Data availability}\\
\indent All data and code supporting the findings of this study are available
from the corresponding author upon request.\\

\noindent \textbf{\large Acknowledgments}\\
\indent This work was funded by LightX project under NWO Open Technology Programme (project number 16714). We thank Ing. E.F.J. Overmars for his support with the Nano L 90-100 laser procurement. We thank the members of LightX user committee: Dr. Jana Sonnenschein (Bayer AG), Dr. Bart Zwijnenburg (Nobian B.V.), Dr. Rob Geertman (Janssen Pharmaceutica), Dr. Andreas Sieber (Lonza Group) and Prof. dr. Antoine E. D. M. van der Heijden (TU Delft).\\

\noindent\textbf{\large Author contributions}\\
\indent N.N., A.V. and H.B.E. designed and performed the experiments;
N.N., R.H. and J.T.P. analyzed the boundary integral element model;
N.N. and H.J.M.K. analyzed the classical nucleation theory;
N.N. and P.J.T.V. analyzed the logistic regression method;
N.N. and V.K. developed the experimental setup;
N.N., R.H., J.T.P. and H.B.E. wrote the manuscript.\\

\noindent\textbf{\large Competing interests}\\
\indent The authors declare no conflict of interest.\\

\bibliographystyle{apsrev4-2}
%\nocite{*}
\bibliography{refs}% Produces the bibliography via BibTeX.

\end{document}

%\preprint{APS/123-QED}

\title{Supplementary Information:\\ Crystal Nucleation Driven by Microjets from Hydrodynamically Interacting Thermocavitation Bubbles
}% Force line breaks with \\
%\thanks{denotes equal contribution}

\author{Nagaraj Nagalingam}
\affiliation{Process \& Energy Department, Delft University of Technology, Leeghwaterstraat 39, 2628 CB Delft, Netherlands}
\author{Adithya Vignesh}
\affiliation{Process \& Energy Department, Delft University of Technology, Leeghwaterstraat 39, 2628 CB Delft, Netherlands}
\author{Vikram Korede}
\affiliation{Process \& Energy Department, Delft University of Technology, Leeghwaterstraat 39, 2628 CB Delft, Netherlands}
\author{Herman J.M. Kramer}
\affiliation{Process \& Energy Department, Delft University of Technology, Leeghwaterstraat 39, 2628 CB Delft, Netherlands}
\author{Peter J.T. Verheijen}
\affiliation{Biotechnology Department, Delft University of Technology, Van der Maasweg 9, 2629 HZ Delft, Netherlands}
\author{Jan Groenewold}
\affiliation{Debye Institute for Nanomaterials Science, Utrecht University, Padualaan 8, 3584 CH Utrecht, Netherlands}
\author{Remco Hartkamp}
\affiliation{Process \& Energy Department, Delft University of Technology, Leeghwaterstraat 39, 2628 CB Delft, Netherlands}
\author{Johan T. Padding}
\affiliation{Process \& Energy Department, Delft University of Technology, Leeghwaterstraat 39, 2628 CB Delft, Netherlands}
\author{Hüseyin Burak Eral}
\email{h.b.eral@tudelft.nl}
\affiliation{Process \& Energy Department, Delft University of Technology, Leeghwaterstraat 39, 2628 CB Delft, Netherlands}

\date{\today}% It is always \today, today,

\maketitle

\tableofcontents

%\section{Optical setup, solution preparation and experimental protocol}

\section{Single bubble}
\begin{figure}[!h]
    \centering
    \includegraphics[width=\linewidth]{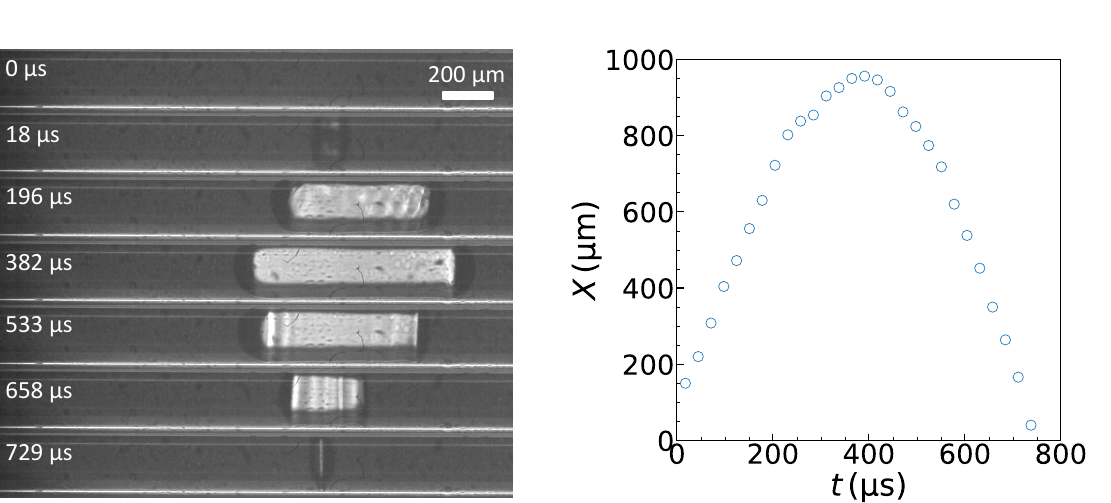}
    \caption{B\textsubscript{1} formation with no B\textsubscript{2}. No crystals were recorded after a fixed observation time of 2\,minutes from laser irradiation. The laser energy supplied is $E_1=304.75\,\upmu$J.}
    \label{single bubble}
\end{figure}

\indent \Cref{single bubble} illustrates B$_1$ formation (with no B$_2$) in a 200\,$\upmu$m capillary. We observed no crystal formation post the bubble collapse within the standard observation time of 2\,minutes. The laser energy supplied is $E_1=304.75\,\upmu$J, which is approximately twice the laser energy supplied for B$_1$ in $200\,\upmu$m capillary in the main text. 

\section{Boundary Integral Element Method (BIEM) solution procedure.}
For an axisymmetric problem with cylindrical coordinates ($r,\theta,z$), all the variables are independent of the rotation angle and therefore can be expressed as \cite{taib1983,taib1985},
\begin{equation}
\frac{1}{|p-q|} \tu{d}S = \frac{4r(\xi)}{D} \left[\left(\dvxi{z}\right)^2 + \left(\dvxi{r}\right)^2\right]^\frac{1}{2} K(k) \tu{d}\xi
\end{equation}
and,
\begin{multline}
\pdvn{}\hspace{-0.3em}\left(\frac{1}{|p-q|}\right) \tu{d}S =  \frac{4r(\xi)}{D^3} \biggl\{\biggl[\dvxi{z}(r(\xi)+r_i) - \dvxi{r}(z(\xi)-z_i) \\
- \frac{2}{k^2(\xi)}\dvxi{z}r_i\biggl] \frac{E(k)}{1-k^2(\xi)} + \frac{2}{k^2(\xi)}\dvxi{z}r_iK(k) \biggl\} \tu{d}\xi.
\end{multline}
With each elements of the boundary parameterized by $\xi \in [0,1]$, $p=(r_i,z_i)$ and  $q=(r(\xi),z(\xi))$. $D=[(r(\xi)+r_i)^2+(z(\xi)-z_i)^2]^\frac{1}{2}$ and $k = 2\sqrt{r(\xi)r_i}/D$. $K(k)$ and $E(k)$ are the complete elliptic integrals of the first and second kind respectively, approximated by the polynomial expansion \cite{hastings2015},
\begin{gather}
K(k) = P(\eta) - Q(\eta) \ln(\eta)\\
E(k) = R(\eta) - S(\eta) \ln(\eta).
\end{gather}
$P$, $Q$, $R$ and $S$ are tabulated polynomials of fourth-order, and $\eta = 1-k^2$.

When the point under consideration ($p$) lies within the interval of integration, logarithmic singularity in $K(k)$ and $E(k)$ exists at $\eta=0$. In addition, a special case exists when this point lies over the axisymmetric axis. The detailed calculations required for these cases are reported in thesis by Taib \cite{taib1985}.\\

\noindent \textbf{Numerical integration} \\
\indent The integral equation is solved numerically along the surface $\Gamma$ as \cite{taib1983,10.1016/j.enganabound.2012.04.009}, 
\begin{multline}
2\pi \phi_i = \sum_{j=1}^N \int_{S_j} \pdvn{}\hspace{-0.3em}\left(\phi_j\right) \, \frac{1}{|p_i-q_j|} \tu{d}\Gamma \\ - \sum_{j=1}^N \int_{S_j} \phi_j\, \pdvn{}\hspace{-0.3em}\left(\frac{1}{|p_i-q_j|}\right) \tu{d}\Gamma.
\label{bemnum}
\end{multline}

Since we have a deforming bubble, it is necessary to have a continuously differentiable surface at each node points. For this purpose, the surface contour is approximated by a finite number of cubic elements as \cite{bartels1995},
\begin{gather}
    r(\xi) = a_j^r + b_j^r \xi + c_j^r \xi^2 + d_j^r \xi^3,\\
    z(\xi) = a_j^z + b_j^z \xi + c_j^z \xi^2 + d_j^z \xi^3.
\end{gather}

As either $\phi$ or $\partial\phi/\partial n$ is unknown over the surface, both the parameters are chosen to linearly vary within each surface element as,
\begin{gather}
    \phi(\xi) = [(1-\xi)\hspace{0.3em} \xi] \left[\begin{matrix}
        \phi_j \\ \phi_{j+1} \end{matrix}\right]\\
    \pdvn{\phi(\xi)} = [(1-\xi)\hspace{0.3em} \xi] \left[\begin{matrix}
        (\partial\phi/\partial n)_j \\ (\partial\phi/\partial n)_{j+1} \end{matrix}\right].
\end{gather}

For each element, the integrals on the hight-hand side of the \Cref{bemnum} can be evaluated as \cite{10.1029/WR015i005p01102}, 
\begin{equation}
    I_e = [k_1] \left[\begin{matrix}
        \phi_j \\ \phi_{j+1} \end{matrix}\right] - [k_2] \left[\begin{matrix}
        (\partial\phi/\partial n)_j \\ (\partial\phi/\partial n)_{j+1} \end{matrix}\right]
\end{equation}
with
\begin{gather}
    [k_1] = \int_0^1 [(1-\xi)\hspace{0.3em} \xi] \hspace{0.3em} \pdvn{}\hspace{-0.3em}\left(\frac{1}{|p_i-q_j|}\right) \tu{d}\xi  \\
    [k_2 ] = \int_0^1 [(1-\xi)\hspace{0.3em} \xi] \hspace{0.3em} \frac{1}{|p_i-q_j|} \tu{d}\xi. 
\end{gather}

The global equations are obtained by summing the contributions from all elements. Thus
\begin{equation}
    H_{i,j} \phi_j = G_{i,j} \left(\pdvn{\phi}\right)_j, \hspace{0.3em} i,j = 1,2,3,...,N
\end{equation}
where 
\begin{gather}
    H_{i,j} = \sum [(k_1)_{i,j} + \delta_{i,j}\, c(p)], \hspace{0.3em} \tu{where} \\
    \delta_{i,j}=1\,\, \forall\,\, i=j \hspace{0.5em} \tu{and} \hspace{0.5em} \delta_{i,j}=0\,\, \forall\,\, i\neq j,\\
    G_{i,j} = \sum (k_2)_{i,j}.
\end{gather}

\Cref{bemnum} is composed of Fredholm integral equation of the first and second kind. During the time marching process, sawtooth instability over the moving boundary profile may appear due to the numerical solution to Fredholm integral equation of the first kind \cite{10.1016/j.enganabound.2012.04.009}. Therefore a five-point smoothing technique introduced by Longuet-Higgins and Cokelet \cite{Longuet} for equally spaced points is used,
\begin{equation}
    \widehat{f}_i = f_i - \omega (f_{i-2}-4f_{i-1}+6f_i-4f_{i+1}+f_{i+2}).
\end{equation}
Where $\omega=1/16$, $f_i$ is the quantity prior to smoothing and $\widehat{f}_i$ is the smoothed quantity. Besides, the nodes of the moving boundaries after updation can become too close to (or too far from) each other, which may also cause numerical instabilities \cite{10.1007/s00162-009-0123-8}. Therefore, the boundary is re-discretized after each time step.\\

\noindent \textbf{Initialization} \\
\indent The computational domain has the $\tu{B}_1$ interface on one end with the other end open. Gravity is neglected as the tube is horizontal and the vertical dimension (capillary diameter) is $O(100\,\upmu$m). The initial $\tu{B}_1$ interface curvature radius is defined using the contact angle of B$_1$ with the capillary wall, $\theta$. It is to be noted that this contact angle is dynamic during the simulations run and changes based on the fluid flow. The bubble $\tu{B}_2$ with radius $R_\tu{0}$ is positioned at distance $h$ away from the $\tu{B}_1$ interface.

The node that connects the $\tu{B}_1$ interface to the capillary wall is considered a part of both the interface and the capillary wall. While solving the boundary integral equation, we treat this common node as a part of wall and impose $\partial\phi/\partial n = 0$ on it. The velocity of this common node along the wall ($\partial z/ \partial t$) is computed using the gradient $\Delta\phi/\Delta z$ at this node with respect to the neighbouring node on wall.\\

\noindent \textbf{Validation} \\
\indent Below we illustrate the validity of model with respect to the initial contact angle ($\theta$) of $\tu{B}_1$. $v_\tu{j,pk}$ is the maximum velocity a jet can achieve during its evolution - a numerical model is necessary to estimate the temporal change in jet velocity. The analytical expression for $v_\tu{j,pk}$ is proposed by Peters \etal \cite{10.1017/jfm.2013.26} as,
\begin{equation}
    v_\tu{j,pk} \propto \frac{\cos{\theta} (1+\beta \cos{\theta})}{b\log(\frac{\sin{\theta-b}}{1-b}) + \sin{\theta} - 1}
    \label{Eqn theta}
\end{equation}
where $b=1+h_0\cos{\theta}$. $h_0$ is a fitting parameter and $\beta$ is a proportionality factor.
\begin{figure*}[!ht]
    \centering
    \includegraphics[width=0.95\linewidth]{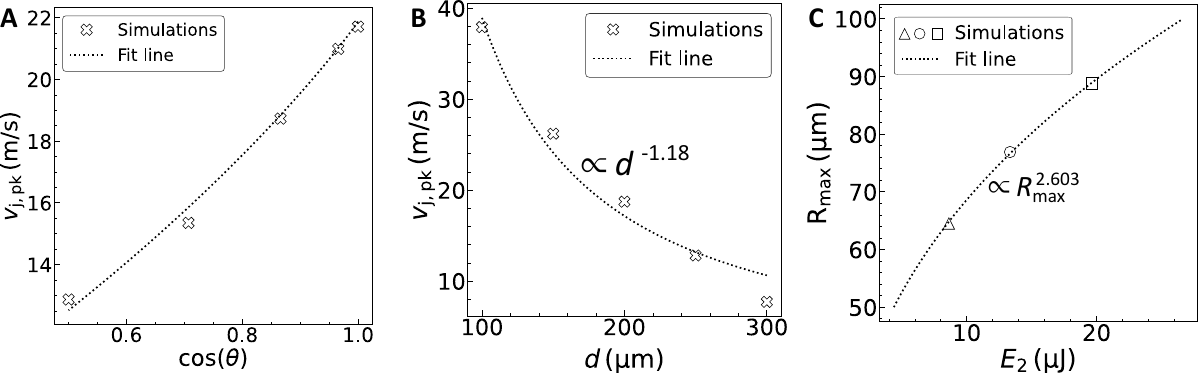}
    \caption{Validation of the Boundary Integral Element Method code. ($A$) Influence of the jet velocity due to change in the initial contact angle. The simulations are for $d=200\,\upmu$m, $h=275\,\upmu$m and $R_\tu{max}=88.8\,\upmu$m. The fit line correspond to \Cref{Eqn theta} with 11.75 as the constant of proportionality. ($B$) Peak jet velocity ($v_\tu{j,pk}$) as a function of capillary diameter ($d$) from simulations. ($C$) Simulated $R_\tu{max}$ as a function of absorbed laser energy in experiments, $E_2$. The average standard error over mean for $E_2$ is $\pm 0.3\,\upmu$J. $R_\tu{max}$ is the maximum radius of the bubble when simulated in an unbound liquid domain - with no capillary walls or $\tu{B}_1$ surface. The three data points correspond to the examples from Fig.\,2$C$ in the main text.}
    \label{validation}
\end{figure*}

\Cref{validation}$A$ shows the relation between the peak jet velocity ($v_\tu{j,pk}$) and the cosine of contact angle. The fit line is derived from \Cref{Eqn theta} with 11.75 as the constant of proportionality and by substituting $h_0=0.26$ and $\beta=0.84$ for $d=200\,\upmu$m \cite{10.1017/jfm.2013.26}. The model thus predicts the correlation same as the analytical expression from literature. In addition, the change in $v_\tu{j,pk}$ for $\zeta=100-500$ was less than $5\%$ relative to $\zeta=313$.

In \Cref{validation}$B$ the model predicts a power law dependence on $d$ for $v_\tu{j,pk}$ with $-1.18$ index, consistent with the previous observations \cite{10.1103/PhysRevX.2.031002}. \Cref{validation}$C$ shows that the $R_\tu{max}$ from simulations to directly correlate to the laser energy from experiments. The determined power law fit
index $E_2 \propto R_\tu{max}^{2.603}$ is consistent with the reported empirical relation for energy difference above a threshold laser energy \cite{10.1103/PhysRevLett.88.078103}.\\ 

\noindent \textbf{Analytical equation for $\zeta$ calculation}\\
\indent With the dimensionless initial radius of the bubble $\widehat{R}_0=R_0/R_\tu{max}$, we calculate \cite{10.1016/j.enganabound.2012.04.009}
\begin{equation}
    \zeta = \left[ \widehat{R}_0^3 - 1 - \frac{3}{\alpha} (1-\widehat{R}_0^2) \right] \frac{(\lambda-1)}{\widehat{R}_0^{3\lambda} (1-\widehat{R}_0^{3(1-\lambda)})}.
    \label{zeta}
\end{equation}
The above relation is derived from the Rayleigh–Plesset equation, using the boundary conditions $\tu{d}R/\tu{d}t=0$ at $R=R_0$ and $R=R_\tu{max}$. Thus Eq.\,\ref{zeta} can be used to calculate $\zeta$, which is used as one of the initial conditions provided we establish $R_0/R_\tu{max}$ and $\lambda$.

\section{The crystal count for experiments reported in Fig.\,3 of main text}
\begin{figure*}[!th]
    \centering
    \includegraphics[width=0.95\linewidth]{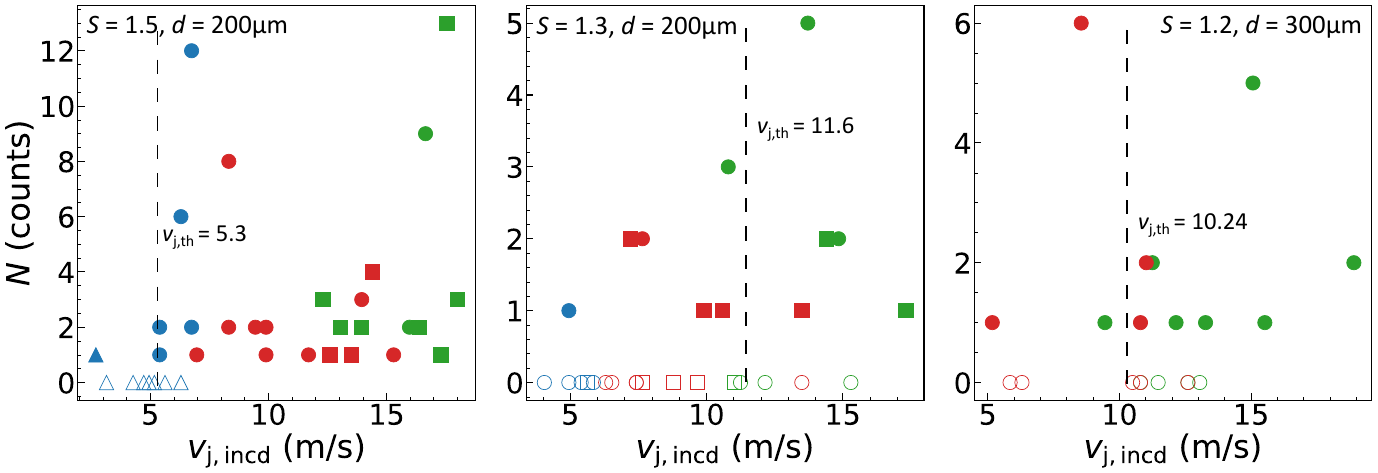}
    \caption{Crystal count. The recorded crystal count for experiments from Fig.\,3 of main text. The markers are adapted from Fig.\,3 in main text. The crystals were recorded after a fixed observation time of 2\,minutes from laser irradiation.}
    \label{crystal count}
\end{figure*}

\indent The crystal count recorded in experiments are reported in \Cref{crystal count}. The results are for the experiments reported in Fig.\,3 of the main text.

\section{Bubble pair experiments within ${300\,\upmu \tu{m}}$ capillary}
\begin{figure}[!h]
    \centering
    \includegraphics[width=\linewidth]{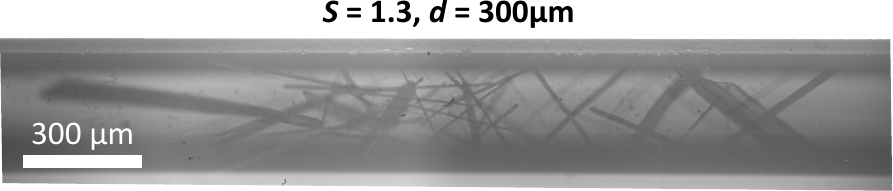}
    \caption{Illustration of crystal formation. The crystals were recorded after a fixed observation time of 2\,minutes from laser irradiation.}
    \label{crystal count in 300}
\end{figure}

\indent \Cref{crystal count in 300} illustrates the crystal formation in experiments for $S=1.3$ within $300\,\upmu$m capillary. The experiments involved formation of bubble pairs with $E_1=219\,\upmu$J and $E_2=16.9\,\upmu$J, similar to the experiments from the main text. In total the experiment was repeated 8 times out of which 5 trials resulted in $>20$ crystals and the other 3 trails resulted in $2-3$ crystals.

\section{Index $n$ values from literature}
\Cref{n value} contains the index value $n$ from the relation $J\propto\dot{\gamma}^n$ (in main text) referenced from literature.
\begin{table*}[!ht]
  %\small
  \centering
  \caption{Empirically observed $n$ values from literature. $\dot{\gamma}$ is the shear rate. The induction time is for the highest shear rate reported in the respective literature.}
    \begin{tabular}{>{\centering\arraybackslash}p{4em}>{\centering\arraybackslash}p{11em} >{\centering\arraybackslash}p{5em} >{\centering\arraybackslash}p{5em} >{\centering\arraybackslash}p{25em}} \hline
    \textbf{n} & \textbf{Compound} & \textbf{$\dot{\gamma}$\,(1/s)} & \textbf{Induction time (s)} & \textbf{Experimental setup design} \\ \hline
    1.443 & lysozyme & $0-10$ & - & Lid driven flow employing a rotating chip \cite{10.1021/acs.cgd.9b01584} \\
    1.226 & butyl paraben & $0-200$ & 1200 & Taylor-Couette flow with rotating inner cylinder \cite{10.1021/cg4007636} \\ 
    3.029 & isotactic polypropylene & $0-8$ & - & Linkam CSS-450 shear stage \cite{10.1016/j.polymer.2016.02.059} \\
    2.08 & paracetamol & $0-200$ & 50 & Rotating parallel plate geometry \cite{10.1039/d3sm00528c} \\
    1 & glycine & $10-1000$ & 100 & Rotating Couette flow and capillary flow devices \cite{10.1021/cg5008878,10.1021/acs.cgd.5b01042} \\
    \hline
    \end{tabular}%
    \label{n value}
\end{table*}%

\section{Solute-solution interfacial tension calculation}
\label{Solute-solution interfacial tension calculation}
\Cref{physical parameters - KMnO4} provides the values required for calculation of solute-solution interfacial tension $\sigma_\tu{sl}$ with $B=16\pi v_0^2 \sigma_\tu{sl}^3/(3k_\tu{B}^3T^3)$.

Mersmann proposed a simple relation for the solute-solution interfacial tension calculation from thermodynamics as \cite{10.1016/0022-0248(90)90850-K},
\begin{equation}
    \sigma \propto T \left(\frac{\rho}{M}\right)^{2/3} \log\left({\frac{\rho}{C^*\,\rho_\tu{L}}}\right).
    \label{gamma}
\end{equation}
Where $T$ is the temperature, $\rho$ is the solute density, $M$ molar mass of solute, $\rho_\tu{L}$ is the solution density, and $C^*$ is the weight fraction in g/g of solution. By substituting the respective values from \Cref{physical parameters - KMnO4} and \Cref{physical parameters - KCl} in the above relation, we calculate the $\sigma$ for \ch{KMnO4} relative to $\sigma_\tu{sl}=2.2$\,mJ/m$^2$ for \ch{KCl} \cite{10.22034/ijnc.2019.33448,10.1021/cg8007415}. The calculated value of $\sigma_\tu{sl}$ for \ch{KMnO4} is 3.16\,mJ/m$^2$ at 298\,K.
\begin{table}[!h]
  %\small
  \centering
  \caption{Physical parameter values for aqueous \ch{KMnO4} solution.}
    \begin{tabular}{>{\centering\arraybackslash}p{3em}>{\centering\arraybackslash}p{12.5em} >{\centering\arraybackslash}p{10em}} \hline
    \textbf{Parameter} & \textbf{Description} & \textbf{Value} \\ \hline
    $B$ & Gradient(slope) [Fig.\,4 from main text] & $0.048$\\
    $v_0$ & Solute molecular volume & $9.71\times10^{-29}$ m$^3$\\
    $k_\tu{B}$ & Boltzmann constant & $1.381\times10^{-23}$ m$^2$kg/(s$^2$K)\\ 
    $T$ & Temperature & $298\,\tu{K}$\\
    $\rho_{\tu{KMnO}_4}$ & Solute density & 2700 kg/m$^3$ \\
    $\rho_{\tu{L-KMnO}_4}$ & Solution density & 1060 kg/m$^3$ (at 25\,$^{\circ}$C and $S=1$) \\
    $C^*_{\tu{KMnO}_4}$ & Solute weight fraction & 0.0706 g/g of solution (at 25\,$^{\circ}$C and $S=1$) \\
    $M_{\tu{KMnO}_4}$ & Molar mass of solute & 158.03 g/mol \\
    $N_\tu{A}$ & Avogadro constant & $6.023\times10^{23}$ mol$^{-1}$\\
    \hline
    \end{tabular}%
    \label{physical parameters - KMnO4}
\end{table}%

\begin{table}[!h]
  %\small
  \centering
  \caption{Physical parameter values for aqueous \ch{KCl} solution.}
    \begin{tabular}{>{\centering\arraybackslash}p{3em}>{\centering\arraybackslash}p{12.5em} >{\centering\arraybackslash}p{10em}} \hline
    \textbf{Parameter} & \textbf{Description} & \textbf{Value} \\ \hline
    $v_0$ & Solute molecular volume & $6.238\times10^{-29}$ m$^3$\\
    $\rho_\tu{KCl}$ & Solute density & 1989 kg/m$^3$ \\
    $\rho_\tu{L-KCl}$ & Solution density & 1175 kg/m$^3$ (at 25\,$^{\circ}$C and $S=1$) \\
    $C^*_\tu{KCl}$ & Solute weight fraction & 0.2645 g/g of solution (at 25\,$^{\circ}$C and $S=1$) \\
    $M_\tu{KCl}$ & Molar mass of solute & 74.55 g/mol \\
    \hline
    \end{tabular}%
    \label{physical parameters - KCl}
\end{table}%

\section{Critical cluster size}
\indent Using the free-energy barrier relation for a nucleus to form, the critical cluster radius $r_\tu{c}$ is given by \cite{10.1002/crat.200310070},
\begin{equation}
    r_\tu{c} = \frac{2 \sigma_\tu{sl} M}{8.314 \rho T \ln{S}}.
\end{equation}
In the above equation the $\sigma_\tu{sl}$ calculated in \Cref{Solute-solution interfacial tension calculation} is used. For \ch{KMnO4} the values reported in \Cref{physical parameters - KMnO4}, such as $M$, $\rho$ and $T$ are used for calculation. The minimum value of $S=1.2$ used in this work is used for calculation. The calculation results in $r_\tu{c} = 0.82\,$nm.

\section{Cavity collapse}
\begin{figure}[!th]
    \centering
    \includegraphics[width=\linewidth]{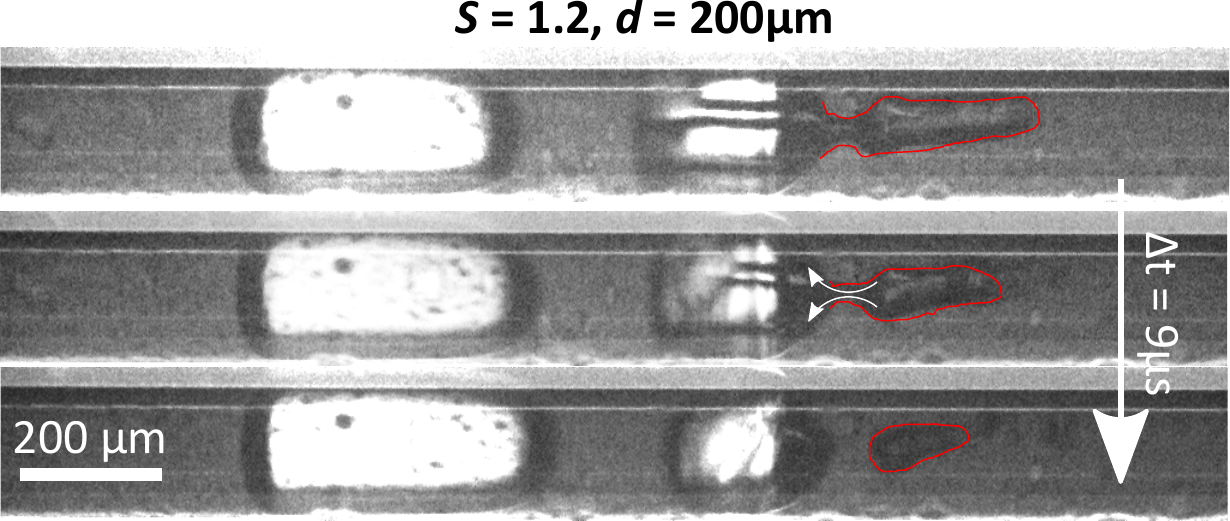}
    \caption{Collapse of the cavity. The red lines mark the boundary of the cavity. The white arrows represent the air exiting the cavity during its collapse.}
    \label{Cavity}
\end{figure}

\Cref{Cavity} shows the collapse of the cavity at the impacted $B_1$ interface by microjet. The collapse of the cavity creates a supersonic airflow. Since velocity is proportional to shear generated, while velocities $O(10\,\tu{m/s})$ create shear rate $O(10^5\,\tu{1/s})$, velocities $O(10000\,\tu{m/s})$ can create shear rate $O(10^8\,\tu{1/s})$.

\section{Videos}
\textbf{Movie S1: Pinch-off jet.} High-speed microscopy experiment with 112,500\,fps showing the first thermocavitation bubble ($B_1$) is triggered at $t=0$. The second bubble $B_2$ is triggered when $B_1$ is at its maximum size thus making the bubbles antiphase. As $B_2$ expands, a jet emerges within the $B_1$ due to hydrodynamic interaction between bubbles (Fig. 1$B$ in main text). Each frame contains a timestamp and scale bar. The confining walls of the microfluidic channel are above and below the viewing plane. The experiment is for $S=1.5$ and $d=200\,\upmu$m, with $\tu{B}_2$ laser energies, $E_2=8.6\,\upmu$J, and for $\tu{B}_1$ laser energy, $E_1 = 157.6\,\upmu$J.\\

\textbf{Movie S2: Straight jet.} High-speed microscopy experiment with 112,500\,fps showing the first thermocavitation bubble ($B_1$) is triggered at $t=0$. The second bubble $B_2$ is triggered when $B_1$ is at its maximum size thus making the bubbles antiphase. As $B_2$ expands, a jet emerges within the $B_1$ due to hydrodynamic interaction between bubbles (Fig. 1$B$ in main text). Each frame contains a timestamp and scale bar. The confining walls of the microfluidic channel are above and below the viewing plane. The experiment is for $S=1.5$ and $d=200\,\upmu$m, with $\tu{B}_2$ laser energies, $E_2=13.3\,\upmu$J, and for $\tu{B}_1$ laser energy, $E_1 = 157.6\,\upmu$J.\\

\textbf{Movie S3: Crystal emergence.} Low-speed microscopy experiment with 3\,fps showing the emergence of \ch{KMnO4} crystals corresponding to the high-speed microscopy experiment from Movie S2: Straight jet. Each frame contains a timestamp and scale bar. The confining walls of the microfluidic channel are above and below the viewing plane. The experiment is for $S=1.5$ and $d=200\,\upmu$m, with $\tu{B}_2$ laser energies, $E_2=13.3\,\upmu$J, and for $\tu{B}_1$ laser energy, $E_1 = 157.6\,\upmu$J.\\

\textbf{Movie S4: Piercing jet.} High-speed microscopy experiment with 112,500\,fps showing the first thermocavitation bubble ($B_1$) is triggered at $t=0$. The second bubble $B_2$ is triggered when $B_1$ is at its maximum size thus making the bubbles antiphase. As $B_2$ expands, a jet emerges within the $B_1$ due to hydrodynamic interaction between bubbles (Fig. 1$B$ in main text). Each frame contains a timestamp and scale bar. The confining walls of the microfluidic channel are above and below the viewing plane. The experiment is for $S=1.5$ and $d=200\,\upmu$m, with $\tu{B}_2$ laser energies, $E_2=19.6\,\upmu$J, and for $\tu{B}_1$ laser energy, $E_1 = 157.6\,\upmu$J.

\textbf{Movie S5: Crystal emergence.} Low-speed microscopy experiment with 3\,fps showing the emergence of \ch{KMnO4} crystals corresponding to the high-speed microscopy experiment from Movie S4: Piercing jet. Each frame contains a timestamp and scale bar. The confining walls of the microfluidic channel are above and below the viewing plane. The experiment is for $S=1.5$ and $d=200\,\upmu$m, with $\tu{B}_2$ laser energies, $E_2=19.6\,\upmu$J, and for $\tu{B}_1$ laser energy, $E_1 = 157.6\,\upmu$J.

\bibliographystyle{apsrev4-2}
%\nocite{*}
\bibliography{refs}% Produces the bibliography via BibTeX.